\documentclass[manuscript,screen]{acmart}
\usepackage{enumitem}
\usepackage{float}
\usepackage{algorithm}
\usepackage[noend]{algpseudocode} 
\usepackage{amsmath}
\algrenewcommand\algorithmiccomment[1]{\hfill$\triangleright$~#1}
\AtBeginDocument{%
  }

\setcopyright{acmlicensed}
\copyrightyear{2018}
\acmYear{2018}
\acmDOI{XXXXXXX.XXXXXXX}
\acmConference[Conference acronym 'XX]{Make sure to enter the correct
  conference title from your rights confirmation email}{June 03--05,
  2018}{Woodstock, NY}
\acmISBN{978-1-4503-XXXX-X/2018/06}

\begin{document}

\title[What Makes LLM Agent Simulations \textit{Useful} for Policy Practice?]{What Makes LLM Agent Simulations \textit{Useful} for Policy Practice? An Iterative Design Study in Emergency Preparedness}

\author{Yuxuan Li}
\affiliation{%
  \institution{School of Computer Science, Carnegie Mellon University}
  \city{Pittsburgh}
  \country{United States}
}
\email{yuxuanll@andrew.cmu.edu}

\author{Sauvik Das}
\affiliation{%
  \institution{School of Computer Science, Carnegie Mellon University}
  \city{Pittsburgh}
  \country{United States}
}
\email{sauvik@cmu.edu}

\author{Hirokazu Shirado}
\affiliation{%
  \institution{School of Computer Science, Carnegie Mellon University}
  \city{Pittsburgh}
  \country{United States}
}
\email{shirado@cmu.edu}


\begin{abstract}
Policymakers must often act under conditions of deep uncertainty, such as emergency response, where predicting the specific impacts of a policy apriori is implausible.
Large Language Model (LLM) agent simulations have been proposed as tools to support policymakers under these conditions, yet little is known about how such simulations become useful for real-world policy practice.
To address this gap, we conducted a year-long, stakeholder-engaged design process with a university emergency preparedness team.
Through iterative design cycles, we developed and refined an LLM agent simulation of a large-scale campus gathering, ultimately scaling to 13,000 agents that modeled crowd movement and communication under various emergency scenarios.
Rather than producing predictive forecasts, these simulations supported policy practice by shaping volunteer training, evacuation procedures, and infrastructure planning.
Analyzing these findings, we identify three design process implications for making LLM agent simulations that are useful for policy practice: start from verifiable scenarios to bootstrap trust, use preliminary simulations to elicit tacit domain knowledge, and treat simulation capabilities and policy implementation as co-evolving.
\end{abstract}



\begin{teaserfigure}
  \centering
  \includegraphics[width=0.8\textwidth]{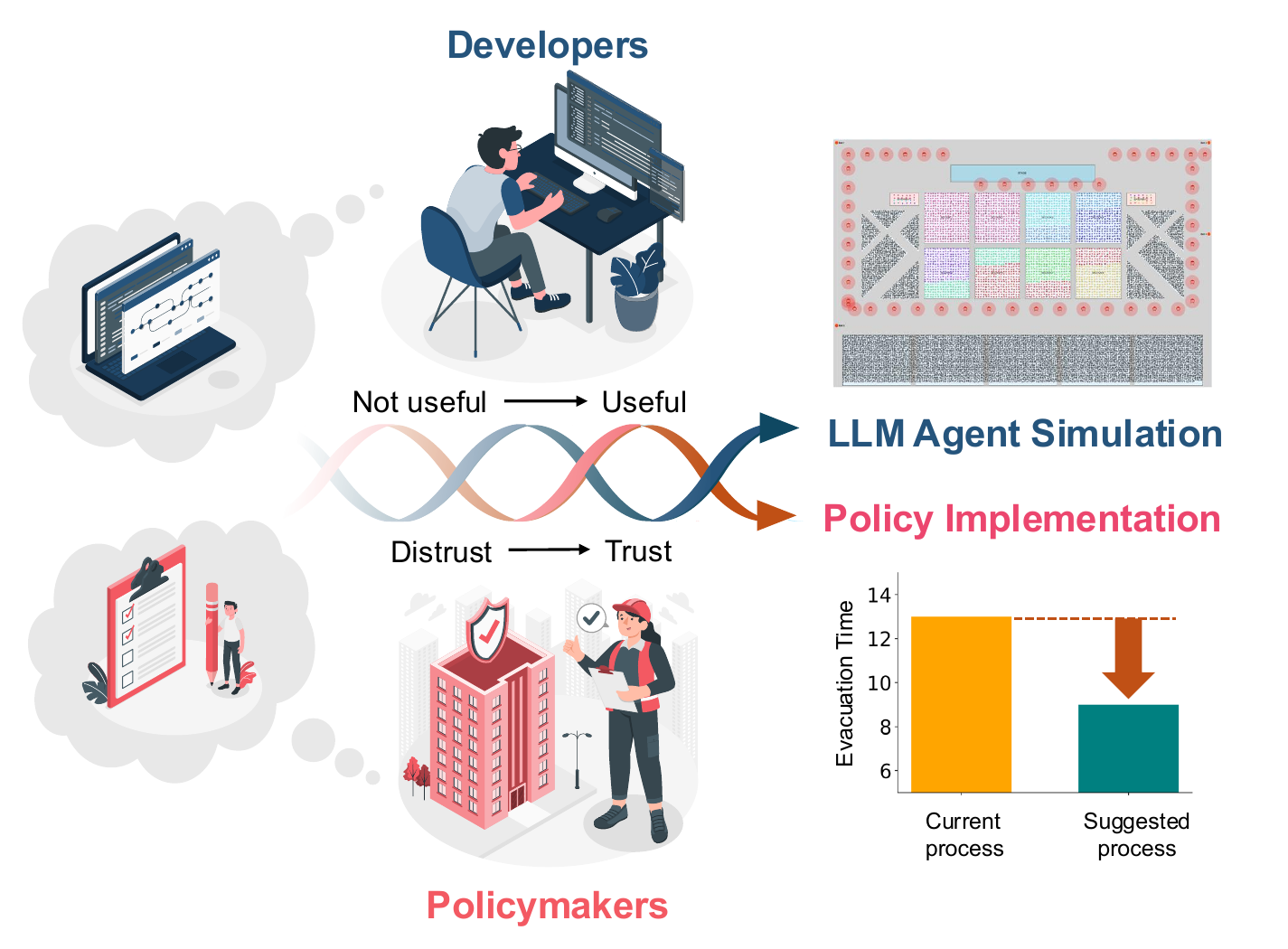}
  \caption{An iterative design engagement with emergency preparedness policymakers developed LLM agent simulations to inform policy implementation.  
  Through cycles of design and validation, simulations shifted from being distrusted to trusted and were ultimately integrated into preparedness practice. 
  This usefulness did not follow a linear path from technical sophistication to institutional adaptation, but instead emerged through an iterative, stakeholder-engaged design process.
  }
  \Description{Diagram showing how developers and policymakers iteratively co-design large language model agent simulations for emergency preparedness. Developers create simulations that policymakers review, moving perceptions from useless to useful and from distrust to trust. A stadium evacuation simulation illustrates how policy changes can be tested, with a bar chart showing improved evacuation time under suggested processes.}
  \label{fig:teaser}
\end{teaserfigure}


\maketitle

\section{Introduction}
Helping policymakers reason and act under conditions of deep uncertainty---situations like emergency response, where outcomes depend on many unknowable environmental and social factors---is a longstanding challenge in decision-making research and human-computer interaction \cite{lempert2002agent, waddell2004case}.
Large language models (LLMs) offer a potential pathway for addressing this challenge: they have been used as interacting agents whose behavior is generated through natural language, enabling large-scale populations to exhibit emergent social behaviors---such as coordination, collective decision-making, and information propagation---grounded in interpretation and communication \cite{park2023generative, sumers2023cognitive, du2023multiagent, piao2025agentsociety, gao2023s3}.
These capabilities have inspired a growing body of work on LLM agent simulation systems, which envision policymakers using these systems to experiment with alternative scenarios, communication strategies, or interventions before implementation \cite{horton2023large, karten2025llm, hwang2025human, hou2025can}.

However, what it means for LLM agent simulations to be useful for policy remains unclear.
Much existing work in such simulations implicitly frames policy in terms of prediction or optimization.
In practice, policy work involves interpreting uncertain information, coordinating across roles, developing procedures, training personnel, and adapting resource allocations under institutional constraints—rather than selecting an optimal action based on forecasts alone \cite{lempert2002agent}.
At the same time, applying LLM agent simulations to policy contexts introduces new risks: they may reproduce stereotypes, amplify biases, or generate outputs that are easy to read but difficult to causally explain or defend within policy practice \cite{arnstein1969ladder, agnew2024illusion, 10.1145/3715275.3732212}.
These concerns foreground the need to understand the conditions under which LLM agent simulations can be meaningfully integrated into policy practice without causing harm.

This motivates a central question for this work: \textbf{How can LLM agent simulations be made genuinely useful for policy practice?}
We address this question by treating policy not as a set of outcomes to be optimized, but as \emph{institutionally situated practice}.
We use the term ``policy'' to refer to both institutionally defined goals (e.g., protecting life and safety) and the concrete procedures, practices, and resource decisions through which those goals are enacted in everyday work.
Accordingly, the challenge is to understand how LLM agent simulations can become useful within the constraints and routines of real-world policy implementation.

Rather than treating simulations as standalone analytical artifacts, the tradition of Human-Computer Interaction (HCI), grounded in participatory and user-centered design methodologies, suggests a design-oriented way forward. 
These approaches emphasize stakeholder-engaged design processes to ensure systems align with actual practices and needs, rather than on abstract technological potential \cite{schuler1993participatory, abras2004user, sanders2008co}.
From this perspective, LLM agent simulations can be studied not only as prediction engines, but as \textit{interactive tools} whose last-mile usefulness depends less on advances in model capability than on how they are iteratively developed with policymakers—grounded in lived challenges and collaboratively explored as part of concrete policy implementation work \cite{lempert2002agent}.

We report on a year-long iterative design engagement with a university emergency preparedness and response team---the institutional decision-makers responsible for preparedness policy implementation (hereafter referred to as the ``policymakers'').
We began with semi-structured interviews to understand existing workflows and decision-making practices, and points where simulation might productively intervene.
Through this process, we identified graduation commencement---a large, complex gathering event involving thousands of students, families, and staff---as a concrete policy domain in which evacuation preparedness was both pressing and open to innovation.
Traditionally, evacuation has primarily been modeled as a physics-based or geometric problem \cite{helbing2005self}.
However, we found that preparedness policy depended critically on deeply social and institutional dimensions --- e.g., communication, coordination, role expectations, and trust --- that are not well captured by these simple physics-based models.
Over five iterative design phases, we developed and refined an LLM agent simulation system to model crowd movement and communication dynamics under varying conditions of scale, physical constraints, and information scarcity.
At each step, we incorporated policymakers’ feedback, validated the simulator against ground-truth data, and worked toward actionable proposals that could inform training, protocols, and the decision-making process itself.

Alongside these design iterations, we analyzed the design process to understand when and why simulations were perceived as relevant, trustworthy, and actionable by policymakers.
From this analysis, we distill three design process implications for making LLM agent simulations useful for policy practice.
First, start with verifiable scenarios to build trust gradually; policymakers engage meaningfully only when their focal aspects of simulation outputs can be checked against reality.
Second, preliminary simulations --- even when imperfect --- can serve as productive technology probes \cite{hutchinson2003technology} that elicit tacit knowledge about roles, relationships, and contextual factors that matter in policy implementation.
Third, simulation and policy development should be treated as interdependent and evolving together; as simulations mature, they reshape policymaker expectations, and as institutional priorities shift, they redefine simulation requirements.

Our contributions are threefold:
\begin{enumerate}
    \item We present an in-depth account of the iterative design of an LLM agent simulation system for emergency preparedness, demonstrating how sustained collaboration enabled the system to evolve from academic prototypes into a tool that informed concrete preparedness practices \footnote{The core contribution of this work is not the system itself, but the design insights we uncovered in using the system as a technology probe.}.
    \item We provide insights into simulation-to-policy relevance by articulating why and how LLM agent simulations come to be perceived as useful, trustworthy, and actionable within institutional policy contexts.
    \item We offer design process implications that provide actionable guidance for researchers and practitioners seeking to develop LLM agent simulations that support policy practice through stakeholder-engaged design. 
\end{enumerate}

Together, these contributions articulate the conditions under which LLM agent simulations can move beyond mere predictive calculators to iterative, stakeholder-engaged tools, and become meaningfully integrated into policy practice.
\section{Related Work}
\subsection{LLM Agent Simulations and Social Modeling}
Simulating human behavior in complex environments has traditionally relied on ABMs \cite{schelling1971dynamic, an2012modeling, macal2005tutorial}.
They have proven effective for modeling physical dynamics, such as crowd flow, geometric bottlenecks, and collision avoidance in evacuation scenarios \cite{joo2013agent, rozo2019modelling, chen2008agent, helbing1995social}.
However, these models often rely on fixed, pre-defined rules that struggle to capture the ambiguity and relational complexity of human decision-making during crises \cite{sun2016simple, deangelis2019decision, kostakos2010brief, aguirre2011contributions}.
Real-world emergencies are rarely reducible to physics problems; they are social events dominated by information scarcity, conflicting social roles, and the need to interpret vague communication.
In addition, accuracies of ABMs often highly depend on the parameters set by their users, making it hard for novices to flexibly tune for new scenarios \cite{groeneveld2017theoretical, ligmann2014using, cheng2018human}.

LLM agents offer a promising alternative by introducing semantic reasoning and generative behavior into social simulation.
Because LLM agents are generative and draw on broad knowledge, they have the potential to address shortcomings of traditional ABMs, including limited ability to model complex human dynamics and heavy dependence on manually tuned parameters \cite{gao2024large, gurcan2024llm}.
Collection of these agents in social contexts have been shown to spontaneously form coalitions, spread invitations, and coordinate collective events \cite{park2023generative, park2024generative}.
Researchers have extended this paradigm to diverse domains, including clinical workflows, classroom dynamics, and macroeconomic trading \cite{binz2025foundation, xie2024can, sumers2023cognitive, Li2023-sg, du2023multiagent}.
In policy-relevant domains, multi-agent systems have been used to explore epidemic response, misinformation spread, and environmental decision-making \cite{hwang2025human, li2024agent, xu2023urban, ma2024computational, hou2025can}.
The appeal of these systems lies in ``in silico'' policy testing---running counterfactual scenarios at scale that would be costly or unethical to attempt in the real world \cite{horton2023large, karten2025llm, li2024large, 10.1145/3706598.3714054}.

Yet, despite their rapid technical progress, these simulations remain almost entirely confined to academic demonstration.
Evaluation has typically emphasized internal qualities such as coherence or plausibility, rather than external markers of usefulness like institutional adoption or decision-making impact.
Policymakers, to our knowledge, have not integrated these systems into practice.

This absence of real-world uptake has fueled a wave of critical concerns.
Some argue that LLM simulations risk reproducing stereotypes or producing ``black-box'' outputs unsuited to accountable governance \cite{arnstein1969ladder, agnew2024illusion, 10.1145/3715275.3732212}, while others warn that by presenting speculative outputs as seemingly concrete, such systems may foster misplaced confidence in decision-makers \cite{kapania2025simulacrum, zhou2024real}.
The result is a widening gap: LLM agent simulations grow more sophisticated in technical fidelity, yet remain disconnected from the institutional realities and trust requirements of policy use.
Our work aims to address this gap by shifting focus from refining agent architectures to designing processes that make simulations useful in institutional policymaking.

\subsection{HCI and Public Policy}
HCI has long been concerned with civic technologies and decision-support systems that shape governance and public policy.
Here in this paper, public policy refers to the processes of implementation through which stable institutional goals are translated into organizational routines, tools, and decisions, rather than to formal rules or legislation alone.
Early work introduced platforms for participatory urban planning and neighborhood decision-making \cite{healey2008civic, abdalla2016decision}. More recent systems like \textit{PolicyCraft} structured deliberations to improve consensus and justification among stakeholders \cite{kuo2025policycraft}.
Studies in crisis informatics highlight similar goals in emergency response, where tools support sensemaking and communication among responders and the public \cite{palen2007citizen, vieweg2010microblogging, starbird2011voluntweeters}.

Across these domains, scholars emphasize the importance of aligning tools with institutional constraints.
For example, Corbett and Le Dantec’s ethnographic co-design with a city immigrant affairs office revealed tensions between efficiency-focused digital tools and the relational work officials valued for building trust \cite{corbett2018going, corbett2021designing, binns2018s}.
Other studies note that legal mandates, organizational hierarchies, and accountability requirements shape how civic or emergency systems can be deployed \cite{binns2018s, alfrink2023contestable, saxena2022unpacking}.
While HCI has articulated a vision of technology that makes policymaking more participatory and evidence-driven \cite{lindblom2018science, vlachokyriakos2016digital, hagan2021prototyping}, recent work find that policy actors themselves remain an underrepresented stakeholder in HCI research \cite{yang2024future}.
This suggests a need for deeper institutional partnerships and design approaches that can bridge between technical innovation and organizational realities \cite{helbing2024co}.

In this study, we build on HCI research that emphasizes aligning technologies with institutional constraints by examining \textit{policy implementation}---defined as the processes through which high-level institutional goals, or \textit{policy} (e.g., protecting life and property in emergencies; see Sec \ref{subsec:coevolution}) are translated into concrete procedures and workflows. 
Through a year-long partnership with emergency preparedness professionals, we investigate how LLM-agent simulations can support institutional processes, and address the practical realities of policy implementation.

\subsection{Engaging Stakeholders in Design}
To navigate such complexities, HCI researchers have developed a wide repertoire of stakeholder engagement methods.
Traditions of participatory design, user-centered design, and community-based co-design have demonstrated how iterative cycles of prototyping, feedback, and negotiation can surface hidden needs and reconcile value tensions \cite{schuler1993participatory, abras2004user, sanders2008co, hutchinson2003technology}.
Longitudinal collaborations with municipal offices \cite{corbett2018going}, community organizations \cite{calvo2017design}, and health institutions \cite{slattery2020research, o2021scoping} underscore the importance of partnerships that adapt alongside institutional contexts.
Agile and iterative approaches have also been adopted in high-stakes domains such as education, health, and emergency response to foster responsiveness and resilience \cite{shore2021art, abrahamsson2017agile, 10.1145/2851581.2892549}.

With the rise of AI-powered systems, however, new challenges emerge.
Unlike traditional tools, AI outputs are probabilistic, opaque, and sometimes unpredictable, complicating stakeholder understanding and co-design \cite{abdul2018trends, Dolata2024DevelopmentITA}.
Recent work has explored participatory approaches to algorithm design to surface values, fairness concerns, and accountability requirements \cite{deng2025weaudit, holstein2019improving}, while others emphasize new representational tools to make AI legible in design workshops \cite{lindley2020researching, ma2020domain}.
Together, these efforts point toward a growing consensus: longitudinal, participatory engagements are critical not only for uncovering requirements, but also for fostering trust, negotiating unpredictability, and embedding AI systems into institutional practice.
This study contributes to these engagement traditions by addressing the specific challenges of LLM agent simulations for emergency preparedness policy implementation, where outputs are both generative and collective, and where contextual requirements are often tacit.
\section{Policy Domain: Emergency Preparedness}

To explore what makes LLM agent simulations useful for policy practice, we identified emergency preparedness as our design context \cite{palen2010vision}.
Emergency preparedness is a particularly fitting domain because field experiments are practically and ethically impossible, observational data are scarce and difficult to generalize, and policymakers must make high-stakes decisions under uncertainty --- where the timing, form, and consequences of emergencies cannot be reliably predicted in advance  \cite{helbing2005self, helbing2006disasters}.
Unlike domains where interventions can be validated through controlled studies, emergency preparedness requires anticipating rare, unpredictable events with limited empirical grounding \cite{comfort2007crisis}.
In practice, policy in emergency preparedness is often enacted through concrete, operational decisions rather than strategic or legislative choices. 
Policymakers must decide how many coordinators or volunteers to deploy, where to position them, which exits to prioritize under different threat scenarios, how to configure accessible seating, and how to frame announcements that guide crowd movement during evacuation. 
These decisions are highly interdependent, context-specific, and they are difficult to evaluate in advance. 
Yet they fundamentally shape how emergency plans are implemented on the ground.
Under such conditions, simulations are less effective when treated as traditional policy analysis tools that prioritize ranking policy actions based on predicted outcomes \cite{morgan1990uncertainty}.
Instead, they can be valued for their ability to support exploration, reflection, and deliberation under uncertainty—helping policymakers examine assumptions, surface blind spots, and reason through possible responses before crises occur \cite{lempert2002agent}.

In addition, many challenges in contemporary emergency preparedness are increasingly shaped by collective communication dynamics rather than individual physical movement alone \cite{castillo2016big, reuter2018fifteen}.
While early research emphasized physics-based movement, it is now understood that physical and social factors interact, with group processes on communication shaping evacuation performance \cite{nilsson2009social, van2021evacuation, templeton2024agent}.
As digital and online technologies have become central to how people receive, interpret, and respond to emergency information \cite{starbird2014rumors, takayasu2015rumor, templeton2023and}, policy effectiveness depends not only on evacuation routes or crowd flows, but also on how messages are framed, how trust is established, and how individuals coordinate in response to uncertain and evolving information \cite{thompson2017evacuation, shirado2020collective}.
These dynamics are inherently social and semantic, involving language, interpretation, emotion, and role expectations that unfold through interaction \cite{drabek2012human, utz2013crisis, jia2021triadic}.
This shift complicates emergency preparedness for conventional policy practices and reduces the utility of tools that assume static or physics-based human behavior.
At the same time, it makes emergency preparedness a particularly relevant context for examining LLM-based agent simulations, which can represent communicative behavior and meaning-making processes alongside physical action.

Given this unpredictability and domain complexity, the usefulness of LLM agent simulations for emergency preparedness can hinge less on advances in modeling fidelity or computational scale, and more on the design process. 
In practice, simulations can \textit{become} useful when they are iteratively developed with policymakers, grounded in their lived challenges, validated against real-world evidence, and collaboratively explored as tools for shaping concrete policy proposals \cite{moss2002policy}.

To situate this inquiry, we partnered with a university's emergency preparedness and response team.
This team is responsible for developing and implementing emergency preparedness policies across diverse scenarios, including large-scale gatherings such as commencement and orientation, natural disasters such as tornadoes and floods, and high-stakes emergencies such as active shooter incidents.
Their policymaking practice was largely experience-driven and discussion-based, and did not involve the routine use of simulation tools.
For recurring events, the team typically adopted plans from previous years as a baseline, using their collective professional judgment to identify incremental improvements.
At the same time, because emergencies are inherently rare and information-poor, the team operated under substantial uncertainty, balancing institutional policy, safety concerns, and communication clarity while coordinating across stakeholders such as university administration, student groups, local police, and neighboring institutions.

A central tool in this process is the \textit{after-action report}. 
Following each major university-wide event, the team documents preparatory decisions and actions taken during the event, highlighting what worked, what failed, and where improvements are needed.
These reports inform future policy implementation and occasionally lead to institution-wide changes.
In this way, after-action reporting serves as a bridge between situated response and long-term organizational learning, making it a critical site for understanding how simulations might inform policy practice.

We began our partnership by introducing prior work on model-based emergency coordination and exploring how it could be extended with recent advances in LLMs. 
At that stage, the policymaker team was already familiar with both the promise and limitations of simulations in their domain.
For example, while they were aware of existing agent-based models for emergency preparedness, they had not incorporated such tools into their process, citing concerns about their ability to capture the situational detail and institutional realities relevant to their practice. 
At the same time, the team recognized the potential relevance of LLMs' natural language capabilities for addressing their challenges related to emergency communication and social media dynamics. 
Several members expressed interest in generating realistic variations in agent behavior through persona-based prompting, viewing this as a way to capture community diversity and stress-test communication strategies.
Notably, the team made no commitment to adapt simulation outputs into policy implementation, emphasizing that the collaboration was exploratory rather than prescriptive.
\section{Methods} \label{sec:methods}
\subsection{Iterative Design and Participants}

We conducted an iterative design engagement with the policymaker team between May 2024 and August 2025, spanning approximately 16 months.
This collaboration was structured around flexible meetings convened as opportunities and needs emerged, typically every 2–6 weeks during active design phases.
Our goal was not to sample policymakers broadly, but to closely examine how simulation usefulness emerges through sustained engagement with a complete decision-making unit within the organization.
While the broader team included many stakeholders, our design work primarily involved five core members who were directly responsible for preparedness and crisis communication: a Senior Director for Disaster Recovery and Business Continuity Services (\textit{P1}), a Senior Director for Reputation and Issues Management (\textit{P2}), a Director for Social Media (\textit{P3}), and two Emergency Preparedness and Response Specialists (\textit{P4}, \textit{P5}).
Together, these individuals represented both policy-level decision makers and operational specialists, allowing our process to bridge institutional strategy with on-the-ground practice.

Our design process unfolded in overlapping cycles of needs assessment through interviews, scenario exploration, simulation prototyping, validation, and proposed policy changes based on simulation outputs.

\subsection{Simulation System (Final Version)} \label{subsec:system}

\begin{figure}[h]
    \centering
    \includegraphics[width=0.7\textwidth]{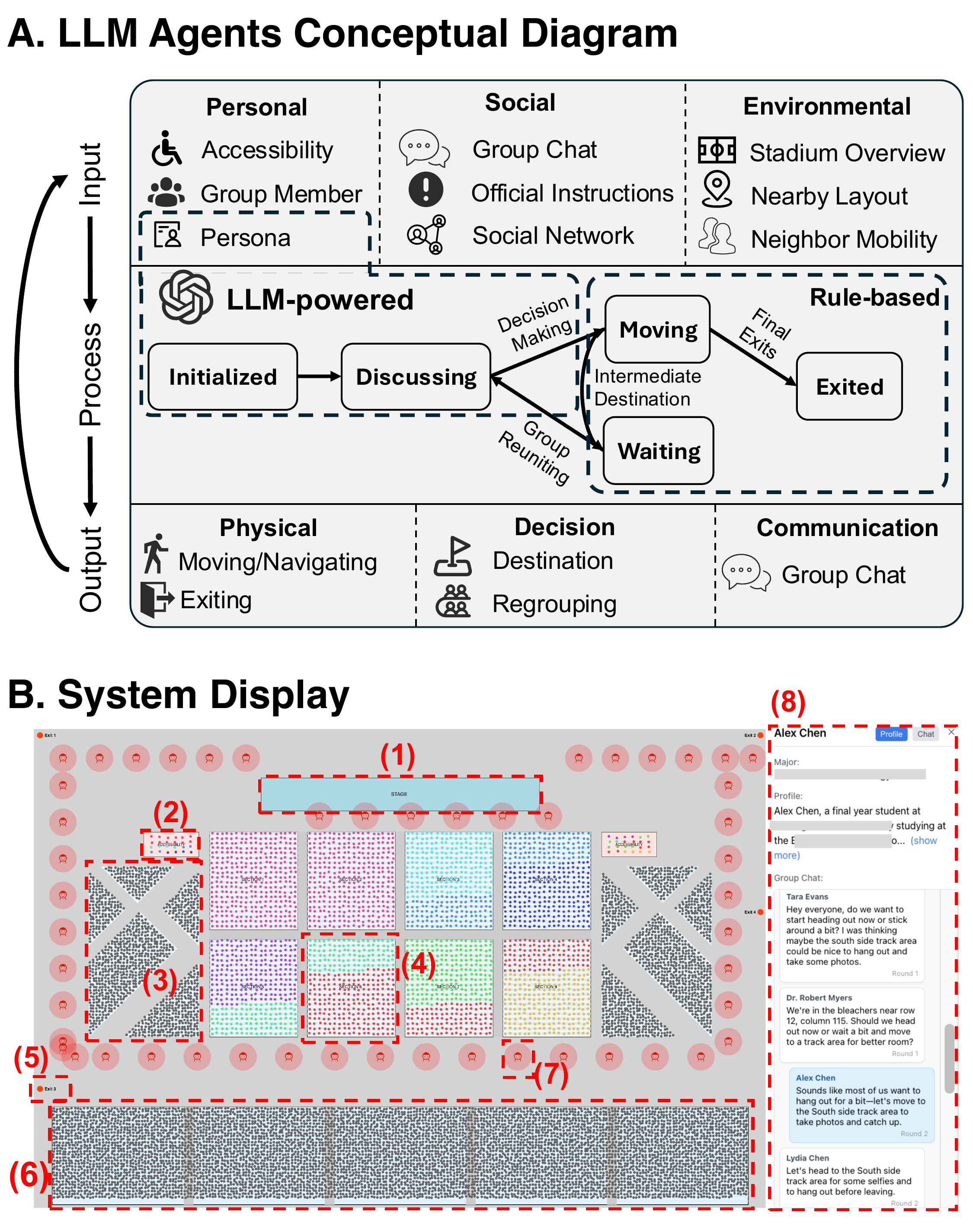}
    \caption{Simulation overview. (A) LLM agents conceptual diagram: Agents receive inputs from three sources: Personal (e.g., accessibility needs, group membership, persona), Social (group chat, official instructions, social network), and Environmental (stadium overview, local layout, neighbor mobility). For each agent, an LLM governs decision-making and communication and hands off to a rule-based controller for embodied action. Outputs include Physical (moving/navigating, exiting), Decision (destination choice, regrouping), and Communication (ongoing group chat). (B) System display: Stadium-scale interface showing physical layouts, agents as colored dots, coordinators, and a per-agent side panel. Numbered callouts: (1) stage area; (2) students with accessibility needs; (3) seating areas for a portion of students' family and friends; (4) seating sections for students from different majors, each major marked by a distinct color; (5) exits; (6) bleacher area with dense family-and-friends seating; (7) coordinators distributed around the field track for directing flow; (8) a per-agent side panel showing persona attributes including name, major and profile, below are group-chat messages between agents that have social relationships and within the same group. Together this system supports simulations with tens of thousands of agents (up to ~13k in our largest runs).}
    \Description{Two-panel figure summarizing our evacuation simulator. Panel A is a block diagram: three input sources---Personal (accessibility needs, group membership, persona), Social (group chat, official instructions, social network), and Environmental (stadium overview, nearby layout, neighbor mobility)---feed each agent’s LLM, which handles decision-making and communication (``Initialized'' to ``Discussing'', and ``Discussing'' to ``Moving'') before handing off to a rule-based controller for embodied action (``Moving'', ``Waiting,'', ``Exited''). Outputs include Physical (moving and exiting), Decision (destination choice and regrouping), and Communication (group chat). Panel B is a stadium-scale display with agents rendered as dots; numbered markers indicate (1) the stage area, (2) a group of students with accessibility needs, (3) seating for some family and friends, (4) student seating sections by major (distinct colors denote different majors), (5) perimeter exits, (6) a dense bleacher area for family and friends, (7) coordinator posts around the track, and (8) a per-agent side panel showing persona fields (name, major, profile) and live group-chat messages among socially connected group members. The system supports large runs with up to approximately 13,000 agents.}
    \label{fig:system}
\end{figure}

Across the design process, we developed and iteratively refined an LLM-based agent simulation system. 
We note that the system itself is \textit{not} the primary contribution of this work; nevertheless, we describe its final form to clarify the artifact through which policymakers engaged with, evaluated, and ultimately adopted simulation outputs.

Figure \ref{fig:system} shows the final system, which simulates approximately 13,000 agents, corresponding to the scale of the university’s commencement event. 
Agents represent students, family members, and emergency coordinators, each with distinct roles, goals, mobility constraints, and social relationships. 
Agents receive inputs from official announcements, local environmental conditions, nearby crowd dynamics, and group communication through online networks. 
OpenAI's GPT-4.1 governs agents’ decision-making and message generation, while a rule-based controller handles physical movement and collision-aware navigation.

The simulation environment models the actual stadium layout, including seating sections, aisles, concourses, exits, and accessibility routes. 
This combination enables the system to represent both physical crowd dynamics and socially mediated behaviors, such as regrouping with family members and responding to coordinator guidance. 
Across scenarios, agents exhibited interpretable context-sensitive behavior: for example, they moved more urgently when informed of approaching severe weather than during routine event dispersal, and responded heterogeneously to localized bomb-threat announcements, with some agents prioritizing rapid exit while others attempted to regroup with family members or parallelized with expressing fear and uncertainty.

Policymakers primarily interacted with the system through visualizations, recorded scenario replays, and comparative scenario outputs rather than direct parameter tuning.
Implementation details, including agent prompting and system architecture, are provided in the Appendix (Sec~\ref{appendix:system_description_final_iteration}). 

In this paper, we treat the system as a technology probe \cite{hutchinson2003technology} --- sufficiently realistic to support validation , deliberation, and preference elicitation --- through how simulation usefulness emerged in practice.
Thus, in the following sections, we show and analyze how usefulness emerged through engagement with this artifact, rather than evaluating the system itself.

\subsection{Data Collection and Analysis} \label{subsec:data_collection_and_analysis}

We collected multiple forms of data across the year-long engagement, including field notes from design and policy implementation proposal sessions, transcripts of meetings, simulation logs and iteration logs from the system, semi-structured interviews, recordings of the commencement event, and relevant email correspondence. 
All data collection procedures were approved by the university’s Institutional Review Board (IRB).

We analyzed our qualitative data as follows.
The first author transcribed and coded meeting transcripts, interviews, recordings of policy implementation proposal sessions, and relevant email exchanges. 
Two other researchers independently reviewed the coded transcripts and provided feedback. 
The first author began by open coding one transcript, then discussed the emerging codes with the other coders to align on coding granularity and analytic focus. 
We then applied open coding across the full set of transcripts and emails, capturing both immediate design concerns and broader reflections about policy relevance \cite{strauss1987qualitative}. 
Together, we refined the codes into higher-level categories and conducted a thematic analysis to surface recurrent patterns \cite{braun2006using}. 
We resolved disagreements through discussion until consensus was reached. 
These qualitative analyses provided the foundation for the detailed accounts of our design iterations and findings presented in the following sections.
\section{Design Iterations} \label{sec:design_iterations}
\begin{figure}[h]
    \centering
    \includegraphics[width=\textwidth]{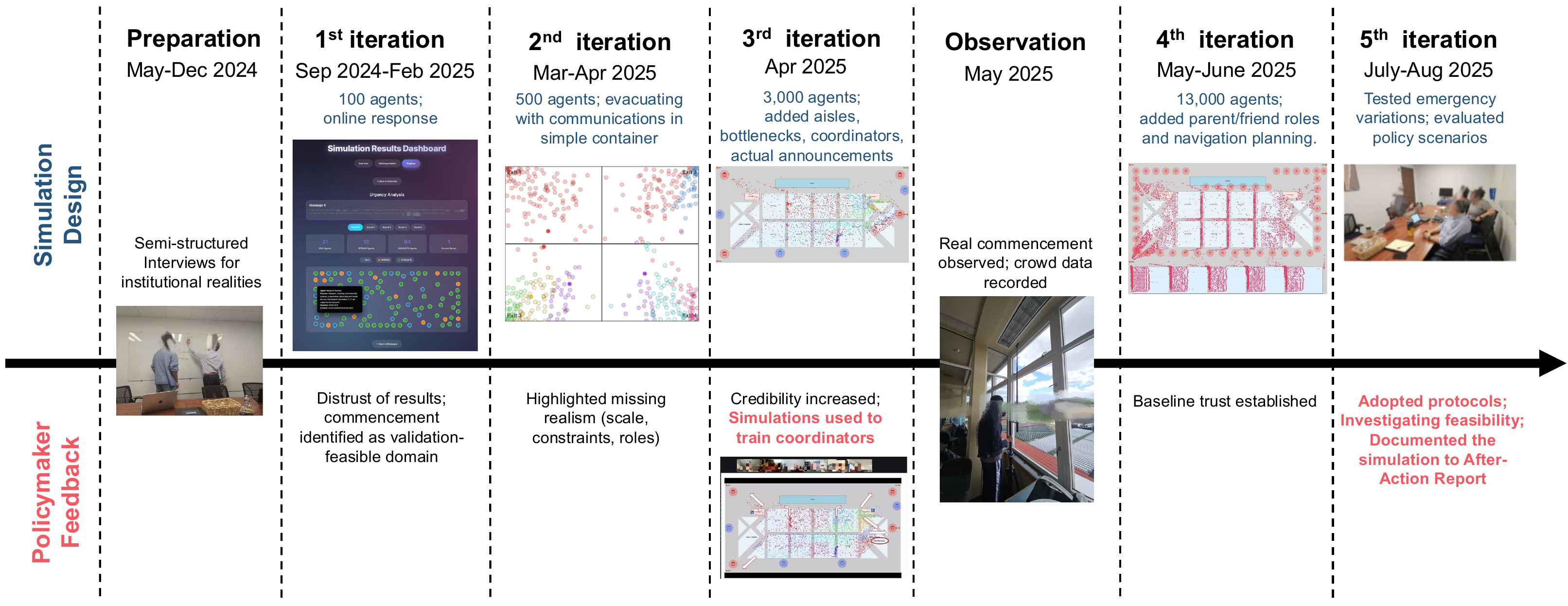}
    \caption{Iterative design process. Across 16 months (May 2024–Aug 2025), we progressed from preparation interviews through five simulation iterations and an in-situ observation. Each phase introduced greater scale and realism (from 100 to 13,000 agents, adding roles, bottlenecks, and family dynamics) and was shaped by policymakers’ feedback. The core system used in later studies (Fig. \ref{fig:system}) was developed by the fourth iteration. Early iterations surfaced distrust and missing realism, later iterations built credibility and training value, and by the final iteration, simulations informed adopted protocols, feasibility assessments, and official after-action reports.}
    \Description{A timeline showing the design process from May 2024 to August 2025. It includes preparation interviews, five iterations of simulations that grow from 100 to 13,000 agents, and an observation at a real commencement. Each stage shows how policymakers’ feedback shaped the next iteration, leading to greater realism, trust, and eventual adoption in training and protocols.}
    \label{fig:timeline}
\end{figure}

We present the iterations in detail to show how simulation usefulness emerged through interaction with institutional constraints, rather than from technical refinement alone.
Across a 16-month engagement, we developed our simulation system through multiple iterations, each shaped by policymakers' feedback and targeted design challenges (Fig. \ref{fig:timeline}).
While these iterations are specific to one institution, the design challenges encountered such as validation and trust calibration reflect broader challenges in LLM agent simulations for policy use cases.

Each iteration is described only to the extent necessary to show how policymakers’ perceptions, validation criteria, and use cases evolved. Technical details are intentionally minimized.

\subsection{Preparation interviews: Capturing Needs and Institutional Realities}

To ground the design process, we conducted three formative semi-structured interviews with emergency preparedness policymakers (\textit{P1–P3}) during the first five months of the project. 
Interviews focused on preparedness workflows and policy development during emergencies.
Insights were synthesized into stakeholder and process maps that documented key actors, responsibilities, and information flows, and were reviewed with participants. 
Full maps are included in the Appendix (Figs. ~\ref{fig:stakeholder_map}, \ref{fig:process_map_1}, \ref{fig:process_map_2}, and \ref{fig:process_map_3}).

\subsection{First Iteration: Simulating Communication and Misinformation Dynamics} \label{subsec:first_iteration}

Based on preparatory interviews, we identified online communication and misinformation as primary concerns during emergencies.
\textit{P1}, \textit{P2}, and \textit{P3} were particularly interested in how official announcements might be misinterpreted and how such misinterpretations could spread through online networks.

To explore this space, we developed an initial prototype simulating 100 LLM agents interacting online. 
Agents, prompted to behave as students, received official lockdown announcements and shared generated responses through a simplified social network. Personas and lightweight state machines introduced variability in interpretation and propagation (implementation details in Appendix, Sec.~\ref{appendix:system_description_iteration_1}).

When we presented this iteration in October 2025, policymakers expressed skepticism about its practical utility. 
While they agreed that modeling misinformation was an important issue, their concerns centered on \textit{validation}: simulated communication dynamics could not be reliably assessed against real-world behavior. 

These discussions led to a key design shift: rather than pursuing simulations that focused exclusively on communication dynamics, we jointly identified evacuation preparedness for the university’s annual commencement as a more suitable domain: one that was both operationally critical and recurrent, enabling direct observation and validation.

\subsection{Second Iteration: Simulating ``Commencement''}
\label{subsec:second_iteration}

Building on the limitations identified in the first iteration, we shifted focus to evacuation preparedness during the university’s annual commencement. 
Unlike online communication scenarios, commencement provided a recurring, institutionally significant setting in which crowd movement and response patterns could be directly observed.

We developed a simplified evacuation simulation with 500 LLM agents prompted to behave as students attending commencement. 
Agents were placed in an abstracted stadium layout, received an official evacuation announcement, exchanged messages with nearby peers, and selected exits over multiple time steps. 
This iteration deliberately prioritized \textit{observability} over realism, serving as a bridge between the communication-focused prototype and a physically grounded evacuation model.

When reviewed by \textit{P1}, \textit{P4}, and \textit{P5} in early April 2025, policymakers immediately identified gaps that limited policy relevance. 
Their feedback emphasized missing contextual details essential for real evacuation planning, including realistic scale, physical constraints (aisles and bottlenecks), accessibility considerations, volunteer coordinator roles, and the use of realistic announcement language. 

Rather than rejecting the simulation, policymakers treated this iteration as a diagnostic artifact. 
Their critiques clarified which contextual features were necessary for trust and directly motivated the next iteration’s increase in scale, environmental fidelity, and role differentiation.

\subsection{Third Iteration: Simulating the University's Commencement} \label{subsec:third_iteration}

Guided by policymakers' feedback, we substantially increased the realism of the simulation to reflect the operational conditions of the university’s commencement.
We incorporated key physical features they identified—including seating sections, aisles, concourses, and bottlenecks—and increased the simulation scale to 3,000 agents.

Agents reflected heterogeneous attendee roles and constraints.
We introduced differentiated mobility profiles for accessibility needs and coordinator agents positioned throughout the stadium to guide evacuation, mirroring volunteer responsibilities during real emergencies.
Movement speed adapted dynamically to local crowd density, enabling realistic congestion patterns to emerge.

We worked with policymakers to obtain the official evacuation message and focused on severe weather scenarios, which they identified as the most operationally relevant emergency type.
Together, these changes shifted the simulation from an abstract model to one recognizable as their institutional environment.

This iteration marked a clear turning point in engagement.
When \textit{P1}, \textit{P4}, and \textit{P5} reviewed the system in late April 2025, they expressed substantially greater confidence in its credibility and relevance.
Rather than questioning validity, policymakers began using the simulation to reason about concrete planning decisions.
For example, \textit{P1} used the visualization to diagnose where volunteer coordination mattered most, noting that congestion emerged well before crowds reached the exits.
In their words, ``by the time the crowd reaches the exits, it's too late --- they're already bumping into each other,'' whereas the key inefficiency occurred earlier, when students selected distant exits and collided along interior routes.
Seeing these dynamics unfold in the simulation helped them reason about where volunteers should be positioned to intervene effectively.

Policymakers also immediately recognized the value of recorded simulations as training materials.
Rather than relying on static slides, they emphasized that coordinators could ``see where congestion is likely to happen and what to expect,'' making the simulations a concrete way to prepare volunteers for real-world conditions.
Following this review, policymakers integrated them into coordinator training sessions ahead of the 2025 commencement and documented the approach in the official \textit{after-action report}.
This marked the first transition of the simulation system from an exploratory technology probe \cite{hutchinson2003technology} to an institutionally-recognized tool supporting policy implementation.

\subsection{Fourth Iteration: Validating Simulation with Real Commencement}
\label{subsec:fourth_iteration}

Following the third iteration, we observed the university’s actual commencement alongside the policymaker team in May 2025.
We examined the venue layout and recorded crowd movement as attendees exited the stadium under IRB approval.
Although no emergency occurred, these observations provided baseline data for validating evacuation dynamics.

During the event, policymakers highlighted areas they routinely monitored for congestion, helping align our analytic focus with practitioner priorities.
One unanticipated pattern proved especially consequential: the central role of students’ family members.
More than half of attendees were family or friends, and after the ceremony many students sought to reunite with them, producing the most severe congestion near family seating areas.

Our prior simulations had modeled only student agents acting independently.
This mismatch revealed a critical omission.
Evacuation behavior was shaped not only by individual movement but by social regrouping, motivating a redesign that treated family members as distinct agents with different goals and interactions.

To address this gap, we developed the most comprehensive version of the system, scaling to nearly 13{,}000 agents—matching the actual event—and incorporating \textit{student}, \textit{family/friends}, and \textit{coordinator} agents.
Agents were seated in designated areas, grouped socially, and designed to communicate and move jointly.
The stadium layout, exits, and accessibility areas reflected the real venue, enabling realistic congestion and regrouping dynamics (Fig.~\ref{fig:system}).

\begin{figure}[h]
    \centering
    \includegraphics[width=1.0\textwidth]{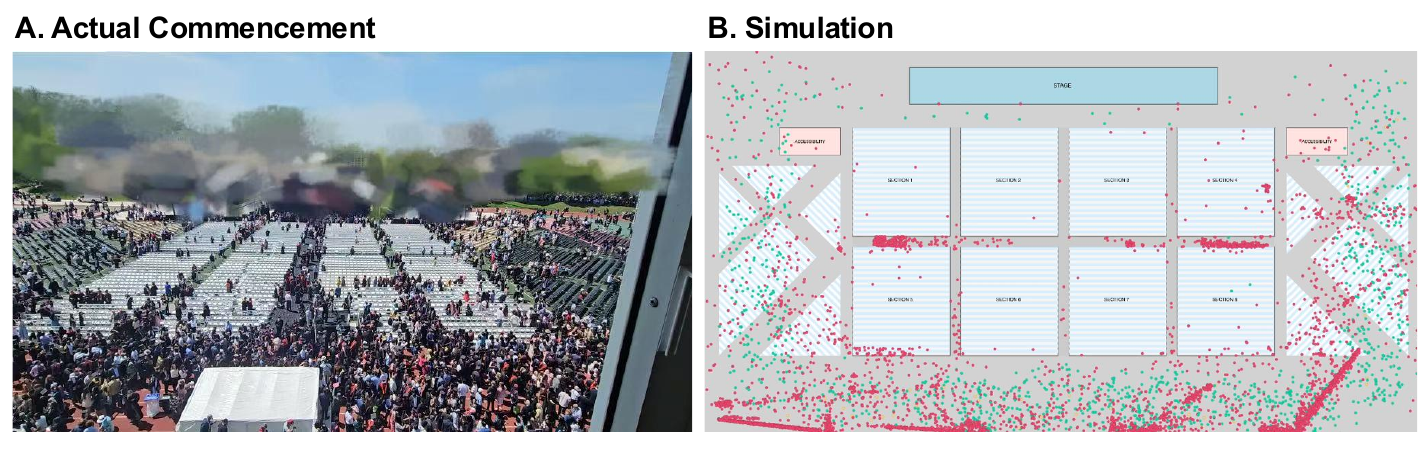}
    \caption{Comparison of crowd movement dynamics between empirical observation and simulation. (A) Photograph of the 2025 commencement, showing attendee movement as the ceremony concluded under routine (non-emergency) conditions. (B) Snapshot from the corresponding simulation, incorporating 13,000 agents with roles for students, family members, and coordinators, represented (colored dots). The simulation (Fig. \ref{fig:system}) reproduced key spatial patterns observed in the event, including congestion in the track areas at the bottom of the frame, supporting alignment between observed and simulated crowd dynamics.}
    \Description{Two side-by-side panels compare crowd movement. The left panel is a photograph of the university’s 2025 commencement, showing large groups of people exiting seating sections and gathering on the track. The right panel is a simulation diagram of the same venue with thousands of small dots representing agents. Both images highlight similar congestion patterns near the track areas close to the camera, illustrating how simulated dynamics align with real-world observations.}
    \label{fig:comparison}
\end{figure}

In June 2025, \textit{P1}, \textit{P4}, and \textit{P5} judged the simulation’s baseline crowd dynamics to align with their routine commencement experience (Fig. \ref{fig:comparison}), which established sufficient trust to explore hypothetical emergencies.

\subsection{Fifth Iteration: Exploring Better Policy Implementation} \label{subsec:fifth_iteration}

In the final iteration, we used the validated simulation to explore concrete policy implementation options raised by policymakers.
We examined operationally plausible variations in emergency type (severe weather and bomb threat), announcement specificity, and coordination strategies, including repositioning coordinators and opening an additional exit.
Interventions were evaluated relative to baseline practices based on ``evacuation efficiency'', measured as the time required for 80\% of agents to evacuate.

This iteration, conducted in August 2025, produced three actionable proposals:
(1) opening a previously unused exit at the northwestern corner of the stadium,
(2) training coordinators to guide both nearby and distant groups during localized threats, and
(3) differentiating evacuation procedures by emergency type and incorporating these distinctions into training materials.

\begin{figure}[h]
    \centering
    \includegraphics[width=0.7\textwidth]{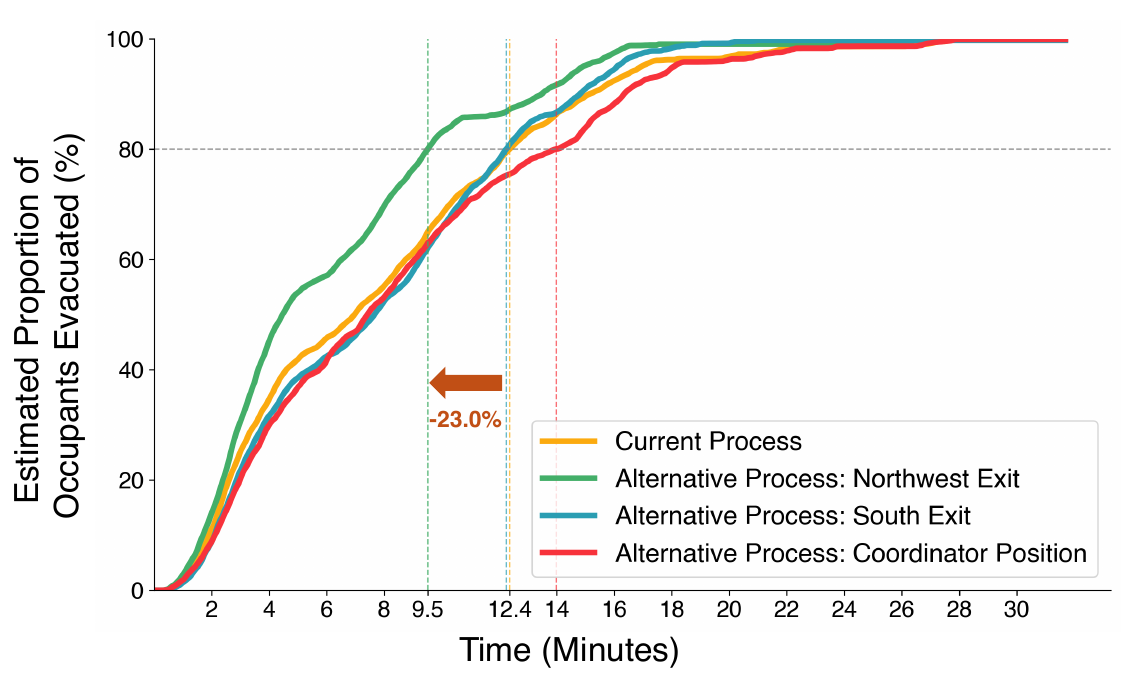}
    \caption{Cumulative evacuation progress across policy alternatives in a severe weather scenario. The figure shows the simulated proportion of agents evacuated over time after an official severe weather announcement instructing evacuation from the stadium. Each curve represents the cumulative percentage of agents who have exited under the current procedure and three alternative strategies explored with the developed stadium system (Fig. \ref{fig:system}). Vertical dashed lines indicate the time required to evacuate 80\% of agents—a threshold that policymakers routinely use to assess evacuation effectiveness. Opening the northwest exit (green) significantly reduces evacuation time relative to the current process (yellow), corresponding to a 23.0\% reduction at the 80\% threshold, while repositioning coordinators or emphasizing the south exit yields only marginal gains. These comparative trajectories supported policymakers’ reasoning about which procedural changes were likely to meaningfully improve evacuation efficiency.}
    \Description{Line chart showing cumulative evacuation over time for four evacuation processes. The x-axis shows time in minutes and the y-axis shows the estimated proportion of occupants evacuated from 0\% to 100\%. Four colored curves represent the current process and three alternative processes: opening a northwest exit, emphasizing a south exit, and changing coordinator positions. All curves rise over time toward full evacuation. Vertical dashed lines mark the time at which 80\% of occupants have evacuated for each process. The northwest exit condition reaches 80\% evacuation earliest, approximately 23\% faster than the current process, while the other alternatives show similar or slightly slower times than the current process.}
    \label{fig:proposal}
\end{figure}

Figure~\ref{fig:proposal} illustrates how simulation outputs informed discussion around Proposal 1.
Across scenarios, variations in coordinators' positions --- which was one of the primary interventions policymakers’ initially considered --- did not substantially affect evacuation time.
In contrast, opening the northwest exit significantly improved evacuation efficiency (23.0\% relative to baseline), whereas other exits produced negligible gains due to local constraints.
Policymakers interpreted the effectiveness of the northwest exit in relation to its proximity to a major congestion area between students and family members.
This intervention was robust in both severe weather and bomb thread scenarios, except when the threat itself was located near that exit.

Following this iteration, the policymakers adopted Proposals 2 and 3, integrating revised coordinator guidance and differentiated emergency protocols into standard operating procedures.
They noted that in man-made threat scenarios such as bomb or active-shooter incidents, operational authority shifts to local police, limiting their ability to fully implement some process changes.
Proposal 1 entered a formal feasibility review, consistent with institutional procedures for infrastructure modifications.
These outcomes were documented in the August 2025 \textit{after-action report}, which committed to continued use of the simulation for preparedness planning and training.
\section{Findings} \label{sec:findings}
We identified five mechanisms through which LLM agent simulations became useful for policy practice in our setting:
\begin{enumerate}
    \item The \textbf{Validation Filter:} Simulations are most useful in scenarios where simulated behaviors can be validated
    \item The \textbf{Trust Bootstrap:} Trust from validating mundane scenarios extends to novel ones
    \item The \textbf{``Fix-It'' Response:} ``Wrong'' simulations surface tacit domain knowledge from policymakers
    \item \textbf{The Details Matter:} Contextual nuance in interaction environments enables policy use
    \item \textbf{Policy-AI Interaction:} Usefulness emerges from co-evolution in policy requirements and simulation capabilities
\end{enumerate}

Across these five mechanisms, policymakers were not primarily evaluating whether the simulator was ``accurate'' in predicting exact motion trajectories.
Instead, they asked a more operational question: \textit{can this simulation be used to justify a concrete change in training, staffing, messaging, or infrastructure?}
In practice, simulation ``usefulness'' emerged when: it created a shared reference point that policymakers could check against something they already knew; it could be interrogated when it looked wrong; and, it could be iterated upon until it supported decisions that policymakers were actually empowered to make.
These findings help explain why the same stochastic and generative affordances that make LLM simulations attractive for exploring ``unknown unknowns'' also make them difficult to adopt without an initial path to validation and trust calibration.
While our findings are grounded in our engagement with just one institution, these mechanisms can reflect recurring conditions of policy-facing simulation work (See Section~\ref{subsec:discussion_1}).
We discuss each in turn below.

\subsection{Validation Filter: Usefulness Requires Scenarios That Can be Validated} \label{subsec:validation_filter}

In our study, simulations were considered useful only when outputs could be validated against policymakers’ experience or ground-truth data—an implicit requirement that became visible through iteration. 
We call this the ``validation filter'': usefulness depends on domains where well-understood scenarios can be jointly checked before extending to hypothetical emergencies. 
In this context, validation did not mean comparing simulation outputs to a single ground-truth value.
Rather, policymakers needed to be able to check simulation behavior against something they already knew from experience, observation, or institutional records.
This surfaced immediately in Iteration 1 (Sec. \ref{subsec:first_iteration}), where \textit{P3} doubted validation was feasible:
\begin{quote}
``It's very hard for us to access social media data in previous emergencies, and some important communication may happen at private Discord channels... And then you have people in other countries who may be having a conversation that's blowing up way beyond what is actually happening here... we don't know, I mean, I don't know how to communicate directly with some of our audiences overseas.''
\end{quote}
Because of this gap, the domain was deemed unsuitable for useful simulation, despite its clear policy relevance. Policymakers instead steered discussions toward scenarios with verifiable outcomes, explicitly welcoming our proposal to observe commencement. As \textit{P1} put it:
\begin{quote}
We will be more than happy to have you here and so what we do is what we actually have an emergency operation center... you have the whole view and you can see actually see the flow of people.
\end{quote}
Commencement was therefore appealing not only because it was high-stakes, but because it recurred and could be directly observed.
This made it possible to establish a shared sense of what ``normal'' looked like before using the simulation to explore more uncertain scenarios.

\subsection{Trust Bootstrap: Trust from Validating Mundane Scenarios Extends to Novel Ones} \label{subsec:trust_bootstrap}

The validation filter creates a paradox: simulations are most valuable for hard-to-observe events \cite{park2022social, park2023generative, binz2025foundation}, yet they are trusted only when grounded in well-understood scenarios.
We found that this paradox could be addressed through a ``trust bootstrap'' process.
In practice, this bootstrapping took the form of a gradual expansion in scope.
Policymakers first accepted the simulator's representation of routine movement patterns and spatial constraints.
Once that foundation was established, they became willing to explore scenarios that layered additional uncertainty on top of this baseline, such as alternative exit strategies or hypothetical disruptions.

In our study, once trust was established through successful validation in mundane scenarios (in our case, attendees exiting the stadium after commencement), that trust could be extended—or ``bootstrapped''—to more novel and speculative scenarios where robust validation was harder or practically impossible.
This process has become especially salient with recent advances in LLMs, which allow simulation systems to adapt to new, natural-language scenarios while building on validated setups.
This dynamic was reflected in \textit{P1}'s comment during the fourth iteration (Sec. \ref{subsec:fourth_iteration}):
\begin{quote}
``...when you showed a baseline evacuation without anything major, it mirrored the 20 minutes of what you all experienced with us with just the regular evacuation. And your simulation mirrored what we experience every time we have it. So that gives us some baseline trust, if you will, to see that it is connecting to what experiences we already had.''
\end{quote}

This baseline trust enabled policymakers to explore more speculative possibilities. 
For example, in our fifth iteration, we demonstrated that opening an additional exit at the northwestern corner could speed up evacuation. 
In response, \textit{P5} proposed another option:
\begin{quote}
What if there was another exit that could potentially be opened?... Right behind it are stairs that lead up to the buildings. I believe the university runs a fence across the back because it’s behind the stadium. Maybe the university could put a temporary gate piece there instead of fencing it off completely... The model (simulation) would be good for this and this is kind of the next step on the maturing the process and considering those additional options.
\end{quote}
We tested the behind-stage exit, but the simulation outputs showed it was less efficient than the northwestern exit.
As a result, the team did not pursue it further.
This underscores how validated trust allowed policymakers to both imagine new options and critically assess their feasibility.

\subsection{``Fix-It'' Response: ``Wrong'' Simulations Surface Tacit Domain Knowledge from Policymakers} \label{subsec:fixit}

We also found that the iterative process was essential because it created opportunities for policymakers to respond to simulations with ``fix-it'' reactions.
Even when policymakers were deeply familiar with their domain, they did not initially know which details were critical for interpreting a simulation until they saw something that looked off.
Incomplete, simplistic, or otherwise ``wrong'' simulations therefore served as productive technology probes \cite{hutchinson2003technology}.
By reacting to outputs that seemed unrealistic or missing, policymakers revealed hidden assumptions and unspoken expertise, which in turn guided refinements that brought the simulations closer to institutional relevance.

For example, our original commencement evacuation simulation did not consider accessibility needs (Sec. \ref{subsec:second_iteration}).
When observing that LLM agents move at the same speed, \textit{P5} immediately flagged this omission:
\begin{quote}
``Did you think about like people with disabilities and they might want to exit like through the one that has like the ramp or anything like that? Did you guys think about those factors?''
\end{quote}
This requirement---accounting for accessibility situations---did not emerge until after policymakers saw a simulation that failed to include it. 

Similarly, while we had incorporated friendship groups based on earlier discussions, the simulation initially modeled only students' evacuation.
After seeing this version, \textit{P4} immediately surfaced a critical missing dimension:
\begin{quote}
``You start thinking about the fact that in a graduation ceremony, for every student, there is potentially minimum of two family members that will be in attendance. And if you're thinking through the concept of a friend group having a very strong connection, you've also got to put into consideration the fact that there is a family connection and they are not with their family in that graduation component.''
\end{quote}
\textit{P5} built on this observation, noting that families would coordinate around shared landmarks: 
\begin{quote}
``Like they'll meet up at landmarks that the family knows... So like if your family's not familiar with CMU and you tell them like let's meet by the goal post that's something visually that they can see.''
\end{quote}
Despite being core to actual evacuation behavior, these family dynamics and landmark-based coordination strategies were never mentioned in initial requirements discussions.
These insights rarely surfaced in interviews and instead emerged when policymakers confronted unrealistic outputs.

\subsection{The Details Matter: Contextual Nuance in Interaction Environments Enables Policy Use}\label{subsec:policy_detail}

LLM agent simulations consist of two core elements: the agents themselves and the environments in which they interact \cite{sumers2023cognitive}. 
In our early iterations, we focused primarily on designing agents and leveraging LLM capabilities. 
However, policymakers consistently pointed out that without sufficient nuance in the interaction environment, the simulations lacked credibility.
When the interaction environment was under-specified, agents could still produce fluent and socially plausible behavior.
However, this surface-level plausibility became a liability rather than an asset: because the interaction environment was under-specified, realistic-looking behavior gave the impression of policy relevance without clearly signaling its limits, making it difficult for policymakers to judge when simulation outputs could or could not inform operational decisions.

In our case for commencement evacuation, these contextual details included the physical (stadium layout, seating), social (family relationships, accessibility needs), procedural (coordinator roles, protocols), and temporal (event sequences, decision timelines).
When these contexts were incorporated, policymakers not only trusted the realism of the outputs but also began to envision practical uses.
For example, in the third iteration, seeing weather-related announcements integrated into the simulation prompted \textit{P4} to imagine how the same system could handle more serious threats:
\begin{quote}
``We were thinking that this evacuation is due to severe weather, which would be a fairly benign reason. But it could also be because of a bomb threat, and that opens up a whole different conversation.'' 
\end{quote}

Similarly, viewing the simulation prompted policymakers to recall physical details they had initially overlooked.
As the simulation became more refined, \textit{P1} noted a missing element:

\begin{quote}
``One of the things that we're not having in the model is the bleachers which is the section in the back, which has a lot more flexibility because we have less volunteers but several thousand people.''
\end{quote}

In academic demonstrations, developers often concentrate on designing LLM agents.
Our findings show that environment design is as consequential as agent design for policy use.
Context reinforced utility by anchoring simulations in recognizable realities, and it sparked imagination about new applications. 
Missing or misaligned contexts could cause temporary dips in confidence, but these gaps also triggered ``fix-it'' responses that revealed additional requirements (see Sec.~\ref{subsec:fixit}).

\subsection{Policy-AI Interaction: Usefulness Emerges From Co-Evolution in Policy Requirements and Simulation Capabilities}\label{subsec:coevolution}

Our iterative process ultimately revealed a co-evolution between policy requirements and simulation capabilities.
While we initially aimed to design simulations that could directly inform ``policy,'' policymakers clarified that our assumption was misplaced. 
\textit{P1} stated: 
\begin{quote}
``These are process changes, not policy or concept of policy is different. Our policy is to implement safe procedures to make sure that we protect life, property. So, policy has a context of what we intend to do... So, our policy is the same. What we're trying to do is to find ways to fulfill the policy, which is protecting life.''
\end{quote}

This distinction matters because policy commitments remain stable while implementation processes adapt to context. 
It also clarified an important boundary on when simulations became actionable.
Policymakers found the simulator most useful when it addressed decisions that fell within their direct authority, such as training practices, coordinator placement, or internal procedures.
In contrast, simulation outputs were harder to act on when potential changes depended on external command structures, legal authority, or approvals beyond the institution.
In these cases, the simulator still supported discussion and sensemaking, but it did not translate as directly into changes in policy implementation.
More broadly, policymakers treated the simulator as a mutual learning device --- useful once baseline trust existed, but always contingent on what institutional knowledge they could supply.
As \textit{P1} later reflected:
\begin{quote}
We are aware that there's so many variables, so we're not going to be going to be able to simulate 100\% the whole. So when we trust the system enough... I think that to mutually learn... Because I think this is for us this is a learning experience. So we know that the system is giving us baseline data, but it's also as good as the information that we provide... to build an own understanding of how the process works.
\end{quote}

Usefulness was therefore recursive: simulations supported policy implementation by helping policymakers refine operational procedures that realize stable policy goals, and those refinements in turn shaped what the simulator needed to represent next.
\section{Discussion}
\subsection{Are LLM Agent Simulations Useful for Policy?}
\label{subsec:discussion_1}

Scholars remain divided on the usefulness of LLM agent simulations for policy.
This debate echoes Bill Buxton's distinction between ``\textit{building the thing right}'' versus ``\textit{building the right thing} \cite{buxton2010sketching}.'' 
While prior work has emphasized increasingly sophisticated agent architectures \cite{park2024generative, sumers2023cognitive, tang2025gensim, gao2023s3}, critics warn that speculative simulations risk being mistaken for predictive truth \cite{arnstein1969ladder, zhou2024real, kapania2025simulacrum}. 

Our engagement suggests that this debate cannot be resolved by evaluating simulation systems in isolation.
Policymakers expressed dissatisfaction with existing tools --- such as tabletop exercises and traditional agent-based models --- that struggle to support reasoning about dynamic, interdependent action.
For example, \textit{P1} explained:
\begin{quote}
``Typically, we’ll sit in a meeting with police, look at a map, and say things like, `I’ll send three officers here.' But we don’t really go deeper into questions like: How long will that take? What are other agents doing at the same time? How do people react dynamically?''
\end{quote}
Policymakers further noted that these tools often abstract away social meaning central to real-world decision-making, resulting in what \textit{P4} described as a ``blank human'' model:
\begin{quote}
``When we normally do tabletop exercises, we’re dealing with a kind of blank human. We don’t know who the person is, and everyone implicitly behaves the same way. But that’s not how real life works.''
\end{quote}
The same limitations also apply to traditional ABMs.
These models can represent evacuation geometry and physical flow more realistically than naive tabletop exercises \cite{joo2013agent, rozo2019modelling, chen2008agent, helbing1995social}, but they still fall short of capturing social and semantic factors central to real-world decision-making, such as communication, social ties, differentiated roles, and heterogeneous interpretations of procedures \cite{sun2016simple, deangelis2019decision, kostakos2010brief, aguirre2011contributions}.
These gaps motivate the exploration of alternative simulation paradigms, not as replacements for existing policy tools, but as complements that foreground communication, interpretation, and institutional roles.

Against this backdrop of institutional dissatisfaction with existing tools, LLM agent simulations offer a different design space: one that integrates language, social meaning, and role-specific reasoning into physically grounded scenarios. 
Our goal is not to claim that LLM agents surpass carefully tuned ABMs in navigation accuracy.
Instead, the persistent challenges that limit the institutional adoption of existing simulation tools motivate exploration of alternative paradigms.
However, LLM agent simulations are not inherently useful; their outputs are stochastic, semantically rich, and difficult to validate, making trust calibration a central design challenge.
We found that LLM agent simulations can \textit{become} useful for policy implementation when embedded in specific design conditions. 
Rather than emerging from predictive accuracy alone, usefulness developed through iterative engagement in which simulations were validated against mundane scenarios, used to surface tacit institutional knowledge, and gradually repurposed for training and operational planning. 
In this role, simulations functioned less as forecasting tools and more as technology probes that supported shared reasoning about complex scenarios.

Nonetheless, caution is warranted. 
As trust builds, model errors and demographic biases may be overlooked \cite{10.1145/3715275.3732212}.  
While such biases can be deliberately leveraged to stress-test policies against real-world behavioral disparities, uncritical reliance on biased outputs risks reinforcing inequities \cite{obermeyer2019dissecting},  requiring a careful distinction between modeling bias and allowing it to shape policy decisions.

These observations also suggest that the usefulness of LLM agent simulations is not uniform across policy domains.
Our findings help identify a set of enabling conditions under which such simulations are most likely to be worth the effort: (1) a recurring baseline scenario that can be observed directly; (2) decision levers that policymakers can realistically change; and (3) outcomes that can be discussed in operational terms rather than abstract social dynamics.

Policy domains that meet these characteristics may similarly support the gradual development of trust in LLM agent simulations.
Beyond emergency preparedness, examples include managing passenger flow and staffing during recurring transit disruptions, coordinating staffing and bed allocation during predictable hospital demand cycles, or planning operations for large, regularly scheduled public events.
In these settings, policymakers can observe baseline conditions, adjust concrete levers such as staffing, routing, or communication workflows, and meaningfully relate simulation outputs to plausible operational changes.

By contrast, domains such as long-term climate adaptation, national security strategy, or online misinformation governance often lack stable, observable baselines and place key decision levers outside the control of any single policymaking unit.
In such contexts, simulations may still offer exploratory value, but translating simulation outputs into actionable institutional change may require forms of validation, interpretation, and organizational alignment beyond those examined in this work.

Rather than arguing that LLM agent simulations should universally guide policy, this work shows how usefulness can emerge through context-sensitive design aligned with institutional practices and constraints.
The central question is therefore not \textit{whether} LLM agent simulations are useful for policy, but \textit{how they can be designed to become useful for policy practice}.

\subsection{How Can LLM Agent Simulations Be Designed to Become Useful for Policy Practice?}

Building on these boundary conditions, we articulate design process implications specific to LLM agent simulations in policy contexts.
While our findings align with established HCI insights that iterative stakeholder engagement increases the likelihood of producing useful technologies, this section highlights three implications that are distinctive to LLM agent simulations when they are intended to support policy implementation rather than exploratory analysis alone.  
Grounded in the five insights identified in Section \ref{sec:findings}, these implications outline how such simulations can be developed and deployed in ways that are genuinely useful within real policy practices. 

\subsubsection{Start with Verifiable Scenarios and Build Trust Gradually}
LLM agent simulations often target complex social phenomena that cannot be directly verified \cite{gao2023s3, hou2025can, karten2025llm}.
Our ``validation filter'' finding (Sec~\ref{subsec:validation_filter}) shows that, in practice simulations only become policymakers could first validate outputs against observable or familiar scenarios.
Because LLM-generated behaviors are inherently stochastic, trust was not established through formal evaluation alone, but through post-hoc alignment with known patterns of behavior.
This unpredictability is also the strength of LLM agent simulations---they allow us to explore unanticipated situations through agent interaction---but it makes the question of validation especially challenging.

Developers and designers should therefore treat validation not as a final evaluation step, but as an early design constraint that shapes which scenarios are explored first.
Starting from routine, verifiable cases allows simulations to establish baseline credibility, which can then be extended to more speculative or unobservable situations.
This staged progression --- what we describe as ``trust bootstrapping'' (Sec~\ref{subsec:trust_bootstrap}) --- enables LLM agent simulations to leverage their generative capacity while remaining usable for policy practice.

\subsubsection{Use Preliminary Simulations to Elicit Tacit Knowledge}
In HCI, prototyping is not just about refinement, but about exploration through progressively increasing fidelity \cite{10.1145/1240624.1240704, buchenau2000experience, hartmann2006reflective, park2022social}.
LLM agent simulations can serve as a similar role in policy contexts; however, our ``fix-it'' response finding (Sec~\ref{subsec:fixit}) highlights an additional benefit.
Because LLM-generated behaviors appear socially plausible, even deliberately preliminary or ``wrong'' simulations can provoke rich critique, surfacing tacit domain knowledge that policymakers often cannot articulate in advance.

This contrasts with typical AI development practices that emphasize accuracy and polished outputs \cite{choudhuri2025guides, brundage2020toward}. 
All simulations are necessarily imperfect---they cannot capture an entire world. 
As Bonini’s paradox reminds us, increasing model complexity does not necessarily increase usefulness \cite{bonini1967simulation}. 
Like a map that is most useful when it selectively simplifies the territory, the challenge in simulation design is not simply to increase fidelity, but to choose carefully which aspects to represent in order to make the model actionable.

Developers should therefore treat interim simulations not simply as flawed prototypes to be corrected, but as probes that invite critique. 
Framing outputs with questions such as ``\textit{what’s missing here?}'' rather than ``\textit{is this correct?}'' helps uncover tacit domain knowledge and institutional practices that can ultimately make the system more useful.
Importantly, the missing elements are not always tied to LLM capabilities themselves, but often to how the broader interaction context is represented. 
Policymakers care about the broader social context, beyond agent behavior alone. 
In our case, much of this feedback concerned the interaction environment---such as spatial layout, family relationships, or coordinator roles---which ultimately proved central to making the simulations useful for policy implementation (Sec~\ref{subsec:policy_detail}). 
These findings underscore that the details surfaced through iterative feedback are not incidental: they are precisely what determines whether a simulation achieves the fidelity needed to inform policy implementation.
At the same time, the ``fix-it'' dynamic can become counterproductive when early simulations generate an unbounded list of realism demands.
In our study, iteration remained productive when critiques could be translated into concrete representational changes and linked to decisions that policymakers could plausibly act on.

\subsubsection{Set a Design Focus on Evolving Simulation Capabilities and Policy Requirements Together}
User-centered design often assumes relatively stable user needs that technology should accommodate \cite{abras2004user, gulliksen2003key, vredenburg2002survey}.
In contrast, we found that simulation capabilities and policymaker requirements co-evolved, with neither fully specified at the outset (Sec~\ref{subsec:coevolution}).
In our case, simulations began as simple demonstrations of evacuation patterns, but over time policymakers used them to explore coordinator placement and differentiated protocols.
At the same time, some simulation-informed ideas were not pursued due to institutional constraints (e.g., law enforcement authority during bomb threats), highlighting that technical possibilities and organizational boundaries must be considered together.
More broadly, simulation insights were easiest to adopt when the same team that interpreted the results could also change the relevant procedures.
Adoption became more difficult when changes required coordination across organizational boundaries or shifts in incident command.

For complex AI systems in institutional contexts, useful design goals cannot be fixed in advance but must be iteratively co-constructed. 
Developers should therefore focus on decision spaces where adaptation is possible—such as training, resource allocation, or communication—while maintaining flexibility for both simulation capabilities and institutional processes to evolve.
This co-evolution enables simulations to move beyond isolated demonstrations and become sustained tools for policy implementation.

\subsection{Limitations and Future Directions}
Our single-domain focus limits generalizability.
Emergency preparedness offers clear success metrics and observable validation opportunities that may not exist in other policy domains.
In addition, our strongest validation opportunities involved routine, non-emergency conditions.
We did not have the chance to observe real-world behavior under active threat, where fear, uncertainty, and authority dynamics may alter coordination in more fundamental ways.
The close collaborative relationships with policymakers may also have influenced outcomes in ways that differ from typical technology adoption scenarios.

Resource requirements raise further questions of scalability.
Our iterative design process required substantial time and commitment from both developers and policymakers---resources many institutions may lack.
User-centered automatic interpretation techniques could help increase both the interpretability and scalability of simulation results \cite{wang2021autods, hase2020evaluating, wright2021recast}.

Ultimately, the usefulness of such tools should be judged by whether suggested changes in policy implementation actually improve safety during emergencies. 
Our study did not directly test implementation results. 
Longitudinal studies are needed to evaluate this question and to trace how simulation tools evolve over extended periods of institutional use. 
A complementary near-term direction is to examine intermediate outcomes that precede rare emergencies, such as whether simulation-based training measurably changes coordinator behavior during drills or improves routine crowd management.
Future research should also test these design conditions across other policy domains and organizational contexts, and explore ways to scale iterative methods to institutions with fewer resources.
Developing evaluation frameworks that center stakeholder experience, rather than traditional accuracy alone, represents another important methodological direction.

Another important direction for future work is comparative evaluation with traditional agent-based models.
LLM agent simulations are a form of agent-based modeling, but differ from traditional approach in how agent behavior is specified: rather than relying on hand-crafted rules or physics-driven dynamics alone, LLM agent simulations use language models to generate semantically rich, context-sensitive behavior.
These differences affect how simulations are designed, validated, and integrated into policy practice.
Accordingly, meaningful evaluation in policy contexts would examine how LLM-based approaches complement --- and trade off with --- the established strengths of traditional agent-based models across dimensions such as interpretability, trust calibration, stakeholder engagement, and alignment with institutional constraints \cite{lempert2002agent}.

Our findings are also constrained by the current capabilities of LLMs---we primarily used OpenAI’s GPT-4o and GPT-4.1. 
As these models advance, new opportunities and challenges for simulation design will likely emerge. 
Nonetheless, our study suggests that building useful LLM agent simulation systems requires shifting focus from technical sophistication toward institutional alignment.
The path from technological demonstration to institutional impact runs not only through better models, but through deep engagement with the organizational and political logics of policymaking itself.

\section{Conclusion}
In this paper, we presented a year-long iterative design engagement with a university emergency preparedness team to examine how LLM agent simulations can become useful for policy practice.
Rather than evaluating simulations as predictive or optimization tools, we traced how participatory design enabled them to evolve from academic demonstrations into institutionally integrated resources for planning, training, and reflection.
Our findings show that usefulness did not arise from technical fidelity alone, but from design practices that gradually built trust, relevance, and shared understanding between developers and policymakers.

By tracing this pathway, we articulated three design process implications for practitioners working at the intersection of HCI, AI, and policymaking: begin with scenarios that admit some form of validation, treat imperfections as opportunities to surface tacit expertise, and co-evolve simulation capabilities with policy implementation needs.
Together, these implications shift attention from asking whether LLM agent simulations are ``accurate enough'' toward understanding how they can be responsibly embedded in real organizational workflows.

As famously observed in statistics, ``\textit{All models are wrong, but some are useful}'' \cite{box1976science}.
Our contribution is to show how LLM agent simulations can be designed to be useful for policy practice, not by claiming predictive certainty, but by aligning technical possibilities with institutional logics, constraints, and decision-making processes. 
More broadly, this work suggests that the value of emerging AI systems in high-stakes domains may hinge less on advances in model capability than on the design processes that connect those capabilities to human judgment, institutional practice, and organizational sensemaking.
We hope this study provides a foundation for future HCI research on designing AI systems that support—not replace—policy practice.
We also hope it informs more careful, grounded integration of LLM-based simulations into real-world institutions.

\section{Acknowledgments}
Portions of the teaser figure (Fig.~\ref{fig:teaser}) are adapted from illustrations by Storyset (https://storyset.com/technology; https://storyset.com/work). We thank Q. Xiao for his insightful advice on the qualitative analysis. This work was supported by the NOMIS foundation and the National Science Foundation under grant \#2316768.

\bibliographystyle{ACM-Reference-Format}
\bibliography{sample-base}

@inproceedings{park2023generative,
  title={Generative agents: Interactive simulacra of human behavior},
  author={Park, Joon Sung and O'Brien, Joseph and Cai, Carrie Jun and Morris, Meredith Ringel and Liang, Percy and Bernstein, Michael S},
  booktitle={Proceedings of the 36th annual acm symposium on user interface software and technology},
  pages={1--22},
  year={2023}
}

@article{park2024generative,
  title={Generative agent simulations of 1,000 people},
  author={Park, Joon Sung and Zou, Carolyn Q and Shaw, Aaron and Hill, Benjamin Mako and Cai, Carrie and Morris, Meredith Ringel and Willer, Robb and Liang, Percy and Bernstein, Michael S},
  journal={arXiv preprint arXiv:2411.10109},
  year={2024}
}

@article{li2024agent,
  title={Agent hospital: A simulacrum of hospital with evolvable medical agents},
  author={Li, Junkai and Lai, Yunghwei and Li, Weitao and Ren, Jingyi and Zhang, Meng and Kang, Xinhui and Wang, Siyu and Li, Peng and Zhang, Ya-Qin and Ma, Weizhi and others},
  journal={arXiv preprint arXiv:2405.02957},
  year={2024}
}

@techreport{horton2023large,
  title={Large language models as simulated economic agents: What can we learn from homo silicus?},
  author={Horton, John J},
  year={2023},
  institution={National Bureau of Economic Research}
}

@article{hou2025can,
  title={Can A Society of Generative Agents Simulate Human Behavior and Inform Public Health Policy? A Case Study on Vaccine Hesitancy},
  author={Hou, Abe Bohan and Du, Hongru and Wang, Yichen and Zhang, Jingyu and Wang, Zixiao and Liang, Paul Pu and Khashabi, Daniel and Gardner, Lauren and He, Tianxing},
  journal={arXiv preprint arXiv:2503.09639},
  year={2025}
}

@article{karten2025llm,
  title={LLM Economist: Large Population Models and Mechanism Design in Multi-Agent Generative Simulacra},
  author={Karten, Seth and Li, Wenzhe and Ding, Zihan and Kleiner, Samuel and Bai, Yu and Jin, Chi},
  journal={arXiv preprint arXiv:2507.15815},
  year={2025}
}

@article{li2024large,
  title={Large language model-driven multi-agent simulation for news diffusion under different network structures},
  author={Li, Xinyi and Xu, Yu and Zhang, Yongfeng and Malthouse, Edward C},
  journal={arXiv preprint arXiv:2410.13909},
  year={2024}
}

@inproceedings{tang2025gensim,
  title={Gensim: A general social simulation platform with large language model based agents},
  author={Tang, Jiakai and Gao, Heyang and Pan, Xuchen and Wang, Lei and Tan, Haoran and Gao, Dawei and Chen, Yushuo and Chen, Xu and Lin, Yankai and Li, Yaliang and others},
  booktitle={Proceedings of the 2025 Conference of the Nations of the Americas Chapter of the Association for Computational Linguistics: Human Language Technologies (System Demonstrations)},
  pages={143--150},
  year={2025}
}

@article{arnstein1969ladder,
  title={A ladder of citizen participation},
  author={Arnstein, Sherry R},
  journal={Journal of the American Institute of planners},
  volume={35},
  number={4},
  pages={216--224},
  year={1969},
  publisher={Taylor \& Francis}
}

@article{zhou2024real,
  title={Is this the real life? is this just fantasy? the misleading success of simulating social interactions with llms},
  author={Zhou, Xuhui and Su, Zhe and Eisape, Tiwalayo and Kim, Hyunwoo and Sap, Maarten},
  journal={arXiv preprint arXiv:2403.05020},
  year={2024}
}

@inproceedings{agnew2024illusion,
  title={The illusion of artificial inclusion},
  author={Agnew, William and Bergman, A Stevie and Chien, Jennifer and D{\'\i}az, Mark and El-Sayed, Seliem and Pittman, Jaylen and Mohamed, Shakir and McKee, Kevin R},
  booktitle={Proceedings of the 2024 CHI Conference on Human Factors in Computing Systems},
  pages={1--12},
  year={2024}
}

@inproceedings{kapania2025simulacrum,
  title={Simulacrum of Stories: Examining Large Language Models as Qualitative Research Participants},
  author={Kapania, Shivani and Agnew, William and Eslami, Motahhare and Heidari, Hoda and Fox, Sarah E},
  booktitle={Proceedings of the 2025 CHI Conference on Human Factors in Computing Systems},
  pages={1--17},
  year={2025}
}

@article{healey2008civic,
  title={Civic engagement, spatial planning and democracy as a way of life civic engagement and the quality of urban places enhancing effective and democratic governance through empowered participation: some critical reflections one humble journey towards planning for a more sustainable Hong Kong: a need to institutionalise civic engagement civic engagement and urban reform in Brazil setting the scene},
  author={Healey, Patsy and Booher, David E and Torfing, Jacob and S{\o}rensen, Eva and Ng, Mee Kam and Peterson, Pedro and Albrechts, Louis},
  journal={Planning Theory \& Practice},
  volume={9},
  number={3},
  pages={379--414},
  year={2008},
  publisher={Taylor \& Francis}
}

@article{abdalla2016decision,
  title={Decision-making tool for participatory urban planning and development: Residents’ preferences of their built environment},
  author={Abdalla, Sahar S and Elariane, Sarah A and El Defrawi, Sarah H},
  journal={Journal of Urban Planning and Development},
  volume={142},
  number={1},
  pages={04015011},
  year={2016},
  publisher={American Society of Civil Engineers}
}

@inproceedings{kuo2025policycraft,
  title={PolicyCraft: Supporting Collaborative and Participatory Policy Design through Case-Grounded Deliberation},
  author={Kuo, Tzu-Sheng and Chen, Quan Ze and Zhang, Amy X and Hsieh, Jane and Zhu, Haiyi and Holstein, Kenneth},
  booktitle={Proceedings of the 2025 CHI Conference on Human Factors in Computing Systems},
  pages={1--24},
  year={2025}
}

@inproceedings{palen2007citizen,
  title={Citizen communications in crisis: anticipating a future of ICT-supported public participation},
  author={Palen, Leysia and Liu, Sophia B},
  booktitle={Proceedings of the SIGCHI conference on Human factors in computing systems},
  pages={727--736},
  year={2007}
}

@inproceedings{vieweg2010microblogging,
  title={Microblogging during two natural hazards events: what twitter may contribute to situational awareness},
  author={Vieweg, Sarah and Hughes, Amanda L and Starbird, Kate and Palen, Leysia},
  booktitle={Proceedings of the SIGCHI conference on human factors in computing systems},
  pages={1079--1088},
  year={2010}
}

@inproceedings{starbird2011voluntweeters,
  title={" Voluntweeters" self-organizing by digital volunteers in times of crisis},
  author={Starbird, Kate and Palen, Leysia},
  booktitle={Proceedings of the SIGCHI conference on human factors in computing systems},
  pages={1071--1080},
  year={2011}
}

@inproceedings{corbett2018going,
  title={Going the distance: Trust work for citizen participation},
  author={Corbett, Eric and Le Dantec, Christopher A},
  booktitle={Proceedings of the 2018 CHI conference on human factors in computing systems},
  pages={1--13},
  year={2018}
}

@inproceedings{corbett2021designing,
  title={Designing civic technology with trust},
  author={Corbett, Eric and Le Dantec, Christopher},
  booktitle={Proceedings of the 2021 CHI Conference on Human Factors in Computing Systems},
  pages={1--17},
  year={2021}
}

@inproceedings{binns2018s,
  title={'It's Reducing a Human Being to a Percentage' Perceptions of Justice in Algorithmic Decisions},
  author={Binns, Reuben and Van Kleek, Max and Veale, Michael and Lyngs, Ulrik and Zhao, Jun and Shadbolt, Nigel},
  booktitle={Proceedings of the 2018 Chi conference on human factors in computing systems},
  pages={1--14},
  year={2018}
}

@inproceedings{alfrink2023contestable,
  title={Contestable camera cars: a speculative design exploration of public AI that is open and responsive to dispute},
  author={Alfrink, Kars and Keller, Ianus and Doorn, Neelke and Kortuem, Gerd},
  booktitle={Proceedings of the 2023 CHI conference on human factors in computing systems},
  pages={1--16},
  year={2023}
}

@inproceedings{saxena2022unpacking,
  title={Unpacking invisible work practices, constraints, and latent power relationships in child welfare through casenote analysis},
  author={Saxena, Devansh and Moon, Seh Young and Shehata, Dahlia and Guha, Shion},
  booktitle={Proceedings of the 2022 CHI Conference on Human Factors in Computing Systems},
  pages={1--22},
  year={2022}
}

@incollection{lindblom2018science,
  title={The science of “muddling through”},
  author={Lindblom, Charles},
  booktitle={Classic readings in urban planning},
  pages={31--40},
  year={2018},
  publisher={Routledge}
}

@inproceedings{vlachokyriakos2016digital,
  title={Digital civics: Citizen empowerment with and through technology},
  author={Vlachokyriakos, Vasillis and Crivellaro, Clara and Le Dantec, Christopher A and Gordon, Eric and Wright, Pete and Olivier, Patrick},
  booktitle={Proceedings of the 2016 CHI conference extended abstracts on human factors in computing systems},
  pages={1096--1099},
  year={2016}
}

@incollection{hagan2021prototyping,
  title={Prototyping for policy},
  author={Hagan, Margaret},
  booktitle={Legal Design},
  pages={9--31},
  year={2021},
  publisher={Edward Elgar Publishing}
}

@inproceedings{yang2024future,
  title={The future of HCI-policy collaboration},
  author={Yang, Qian and Wong, Richmond Y and Jackson, Steven and Junginger, Sabine and Hagan, Margaret D and Gilbert, Thomas and Zimmerman, John},
  booktitle={Proceedings of the 2024 CHI Conference on Human Factors in Computing Systems},
  pages={1--15},
  year={2024}
}

@book{schuler1993participatory,
  title={Participatory design: Principles and practices},
  author={Schuler, Douglas and Namioka, Aki},
  year={1993},
  publisher={CRC press}
}

@article{abras2004user,
  title={User-centered design},
  author={Abras, Chadia and Maloney-Krichmar, Diane and Preece, Jenny and others},
  journal={Bainbridge, W. Encyclopedia of Human-Computer Interaction. Thousand Oaks: Sage Publications},
  volume={37},
  number={4},
  pages={445--456},
  year={2004}
}

@article{sanders2008co,
  title={Co-creation and the new landscapes of design},
  author={Sanders, Elizabeth B-N and Stappers, Pieter Jan},
  journal={Co-design},
  volume={4},
  number={1},
  pages={5--18},
  year={2008},
  publisher={Taylor \& Francis}
}

@inproceedings{10.1145/3715275.3732212,
author = {Li, Yuxuan and Shirado, Hirokazu and Das, Sauvik},
title = {Actions Speak Louder than Words: Agent Decisions Reveal Implicit Biases in Language Models},
year = {2025},
isbn = {9798400714825},
publisher = {Association for Computing Machinery},
address = {New York, NY, USA},
url = {https://doi.org/10.1145/3715275.3732212},
doi = {10.1145/3715275.3732212},
booktitle = {Proceedings of the 2025 ACM Conference on Fairness, Accountability, and Transparency},
pages = {3303–3325},
numpages = {23},
location = {
},
series = {FAccT '25}
}

@inproceedings{10.1145/3706598.3714054,
author = {Jin, Hyoungwook and Yoo, Minju and Park, Jeongeon and Lee, Yokyung and Wang, Xu and Kim, Juho},
title = {TeachTune: Reviewing Pedagogical Agents Against Diverse Student Profiles with Simulated Students},
year = {2025},
isbn = {9798400713941},
publisher = {Association for Computing Machinery},
address = {New York, NY, USA},
url = {https://doi.org/10.1145/3706598.3714054},
doi = {10.1145/3706598.3714054},
booktitle = {Proceedings of the 2025 CHI Conference on Human Factors in Computing Systems},
articleno = {1073},
numpages = {28},
location = {
},
series = {CHI '25}
}

@article{slattery2020research,
  title={Research co-design in health: a rapid overview of reviews},
  author={Slattery, Peter and Saeri, Alexander K and Bragge, Peter},
  journal={Health research policy and systems},
  volume={18},
  number={1},
  pages={17},
  year={2020},
  publisher={Springer}
}

@article{o2021scoping,
  title={A scoping review of the use of co-design methods with culturally and linguistically diverse communities to improve or adapt mental health services},
  author={O’Brien, Jennifer and Fossey, Ellie and Palmer, Victoria J},
  journal={Health \& Social Care in the Community},
  volume={29},
  number={1},
  pages={1--17},
  year={2021},
  publisher={Wiley Online Library}
}

@article{calvo2017design,
  title={Design for social sustainability. A reflection on the role of the physical realm in facilitating community co-design.},
  author={Calvo, Mirian and De Rosa, Annalinda},
  journal={The Design Journal},
  volume={20},
  number={sup1},
  pages={S1705--S1724},
  year={2017},
  publisher={Taylor \& Francis}
}

@book{shore2021art,
  title={The art of agile development},
  author={Shore, James and Warden, Shane},
  year={2021},
  publisher={" O'Reilly Media, Inc."}
}

@article{abrahamsson2017agile,
  title={Agile software development methods: Review and analysis},
  author={Abrahamsson, Pekka and Salo, Outi and Ronkainen, Jussi and Warsta, Juhani},
  journal={arXiv preprint arXiv:1709.08439},
  year={2017}
}

@inproceedings{10.1145/2851581.2892549,
author = {Thomer, Andrea K. and Twidale, Michael B. and Guo, Jinlong and Yoder, Matthew J.},
title = {Co-designing Scientific Software: Hackathons for Participatory Interface Design},
year = {2016},
isbn = {9781450340823},
publisher = {Association for Computing Machinery},
address = {New York, NY, USA},
url = {https://doi.org/10.1145/2851581.2892549},
doi = {10.1145/2851581.2892549},
booktitle = {Proceedings of the 2016 CHI Conference Extended Abstracts on Human Factors in Computing Systems},
pages = {3219–3226},
numpages = {8},
location = {San Jose, California, USA},
series = {CHI EA '16}
}

@inproceedings{abdul2018trends,
  title={Trends and trajectories for explainable, accountable and intelligible systems: An hci research agenda},
  author={Abdul, Ashraf and Vermeulen, Jo and Wang, Danding and Lim, Brian Y and Kankanhalli, Mohan},
  booktitle={Proceedings of the 2018 CHI conference on human factors in computing systems},
  pages={1--18},
  year={2018}
}

@article{Dolata2024DevelopmentITA,
  title={Development in Times of Hype: How Freelancers Explore Generative AI?},
  author={Mateusz Dolata and Norbert Lange and Gerhard Schwabe},
  journal={2024 IEEE/ACM 46th International Conference on Software Engineering (ICSE)},
  year={2024},
  pages={2257-2269},
  url={https://api.semanticscholar.org/CorpusId:266933328}
}

@article{deng2025weaudit,
  title={WeAudit: Scaffolding User Auditors and AI Practitioners in Auditing Generative AI},
  author={Deng, Wesley Hanwen and Claire, Wang and Han, Howard Ziyu and Hong, Jason I and Holstein, Kenneth and Eslami, Motahhare},
  journal={arXiv preprint arXiv:2501.01397},
  year={2025}
}

@inproceedings{holstein2019improving,
  title={Improving fairness in machine learning systems: What do industry practitioners need?},
  author={Holstein, Kenneth and Wortman Vaughan, Jennifer and Daum{\'e} III, Hal and Dudik, Miro and Wallach, Hanna},
  booktitle={Proceedings of the 2019 CHI conference on human factors in computing systems},
  pages={1--16},
  year={2019}
}

@inproceedings{lindley2020researching,
  title={Researching AI legibility through design},
  author={Lindley, Joseph and Akmal, Haider Ali and Pilling, Franziska and Coulton, Paul},
  booktitle={Proceedings of the 2020 CHI Conference on Human Factors in Computing Systems},
  pages={1--13},
  year={2020}
}

@inproceedings{ma2020domain,
  title={How domain experts create conceptual diagrams and implications for tool design},
  author={Ma'ayan, Dor and Ni, Wode and Ye, Katherine and Kulkarni, Chinmay and Sunshine, Joshua},
  booktitle={Proceedings of the 2020 CHI Conference on Human Factors in Computing Systems},
  pages={1--14},
  year={2020}
}

@article{piao2025agentsociety,
  title={AgentSociety: Large-Scale Simulation of LLM-Driven Generative Agents Advances Understanding of Human Behaviors and Society},
  author={Piao, Jinghua and Yan, Yuwei and Zhang, Jun and Li, Nian and Yan, Junbo and Lan, Xiaochong and Lu, Zhihong and Zheng, Zhiheng and Wang, Jing Yi and Zhou, Di and others},
  journal={arXiv preprint arXiv:2502.08691},
  year={2025}
}

@article{gao2023s3,
  title={S3: Social-network simulation system with large language model-empowered agents},
  author={Gao, Chen and Lan, Xiaochong and Lu, Zhihong and Mao, Jinzhu and Piao, Jinghua and Wang, Huandong and Jin, Depeng and Li, Yong},
  journal={arXiv preprint arXiv:2307.14984},
  year={2023}
}

@article{binz2025foundation,
  title={A foundation model to predict and capture human cognition},
  author={Binz, Marcel and Akata, Elif and Bethge, Matthias and Br{\"a}ndle, Franziska and Callaway, Fred and Coda-Forno, Julian and Dayan, Peter and Demircan, Can and Eckstein, Maria K and {\'E}ltet{\H{o}}, No{\'e}mi and others},
  journal={Nature},
  pages={1--8},
  year={2025},
  publisher={Nature Publishing Group UK London}
}

@inproceedings{xie2024can,
  title={Can Large Language Model Agents Simulate Human Trust Behavior?},
  author={Xie, Chengxing and Chen, Canyu and Jia, Feiran and Ye, Ziyu and Lai, Shiyang and Shu, Kai and Gu, Jindong and Bibi, Adel and Hu, Ziniu and Jurgens, David and others},
  booktitle={The Thirty-eighth Annual Conference on Neural Information Processing Systems},
  year={2024}
}

@article{sumers2023cognitive,
  title={Cognitive architectures for language agents},
  author={Sumers, Theodore and Yao, Shunyu and Narasimhan, Karthik and Griffiths, Thomas},
  journal={Transactions on Machine Learning Research},
  year={2023}
}

@INPROCEEDINGS{Li2023-sg,
  title     = "{CAMEL}: Communicative Agents for ``Mind'' Exploration of Large Language Model Society",
  author    = "Li, Guohao and Hammoud, Hasan and Itani, Hani and Khizbullin, Dmitrii and Ghanem, Bernard",
  editor    = "Oh, A and Naumann, T and Globerson, A and Saenko, K and Hardt, M and Levine, S",
  booktitle = "Advances in Neural Information Processing Systems",
  volume    =  36,
  pages     = "51991--52008",
  year      =  2023
}

@inproceedings{du2023multiagent,
  title={Improving Factuality and Reasoning in Multi-agent Debate},
  author={Du, Yilun and Ma, Xueguang and Song, Shuran and Tenenbaum, Joshua B. and Torralba, Antonio},
  booktitle={Forty-first International Conference on Machine Learning (ICML)},
  year={2024}
}

@article{braun2006using,
  title={Using thematic analysis in psychology},
  author={Braun, Virginia and Clarke, Victoria},
  journal={Qualitative research in psychology},
  volume={3},
  number={2},
  pages={77--101},
  year={2006},
  publisher={Taylor \& Francis}
}

@book{strauss1987qualitative,
  title={Qualitative analysis for social scientists},
  author={Strauss, Anselm L},
  year={1987},
  publisher={Cambridge university press}
}

@inproceedings{hwang2025human,
  title={Human Subjects Research in the Age of Generative AI: Opportunities and Challenges of Applying LLM-Simulated Data to HCI Studies},
  author={Hwang, Angel Hsing-Chi and Bernstein, Michael S and Sundar, S Shyam and Zhang, Renwen and Horta Ribeiro, Manoel and Lu, Yingdan and Chang, Serina and Wu, Tongshuang and Yang, Aimei and Williams, Dmitri and others},
  booktitle={Proceedings of the Extended Abstracts of the CHI Conference on Human Factors in Computing Systems},
  pages={1--7},
  year={2025}
}

@article{xu2023urban,
  title={Urban generative intelligence (ugi): A foundational platform for agents in embodied city environment},
  author={Xu, Fengli and Zhang, Jun and Gao, Chen and Feng, Jie and Li, Yong},
  journal={arXiv preprint arXiv:2312.11813},
  year={2023}
}

@article{ma2024computational,
  title={Computational experiments meet large language model based agents: A survey and perspective},
  author={Ma, Qun and Xue, Xiao and Zhou, Deyu and Yu, Xiangning and Liu, Donghua and Zhang, Xuwen and Zhao, Zihan and Shen, Yifan and Ji, Peilin and Li, Juanjuan and others},
  journal={arXiv preprint arXiv:2402.00262},
  year={2024}
}

@book{buxton2010sketching,
  title={Sketching user experiences: getting the design right and the right design},
  author={Buxton, Bill},
  year={2010},
  publisher={Morgan kaufmann}
}

@inproceedings{hutchinson2003technology,
  title={Technology probes: inspiring design for and with families},
  author={Hutchinson, Hilary and Mackay, Wendy and Westerlund, Bo and Bederson, Benjamin B and Druin, Allison and Plaisant, Catherine and Beaudouin-Lafon, Michel and Conversy, St{\'e}phane and Evans, Helen and Hansen, Heiko and others},
  booktitle={Proceedings of the SIGCHI conference on Human factors in computing systems},
  pages={17--24},
  year={2003}
}

@inproceedings{park2022social,
  title={Social simulacra: Creating populated prototypes for social computing systems},
  author={Park, Joon Sung and Popowski, Lindsay and Cai, Carrie and Morris, Meredith Ringel and Liang, Percy and Bernstein, Michael S},
  booktitle={Proceedings of the 35th Annual ACM Symposium on User Interface Software and Technology},
  pages={1--18},
  year={2022}
}

@article{box1976science,
  title={Science and statistics},
  author={Box, George EP},
  journal={Journal of the American Statistical Association},
  volume={71},
  number={356},
  pages={791--799},
  year={1976},
  publisher={Taylor \& Francis}
}

@book{bonini1967simulation,
  title={Simulation of information and decision systems in the firm},
  author={Bonini, Charles P and others},
  year={1967},
  publisher={Markham Chicago}
}

@article{obermeyer2019dissecting,
  title={Dissecting racial bias in an algorithm used to manage the health of populations},
  author={Obermeyer, Ziad and Powers, Brian and Vogeli, Christine and Mullainathan, Sendhil},
  journal={Science},
  volume={366},
  number={6464},
  pages={447--453},
  year={2019},
  publisher={American Association for the Advancement of Science}
}

@article{helbing2005self,
  title={Self-organized pedestrian crowd dynamics: Experiments, simulations, and design solutions},
  author={Helbing, Dirk and Buzna, Lubos and Johansson, Anders and Werner, Torsten},
  journal={Transportation science},
  volume={39},
  number={1},
  pages={1--24},
  year={2005},
  publisher={INFORMS}
}

@incollection{helbing2006disasters,
  title={Disasters as extreme events and the importance of network interactions for disaster response management},
  author={Helbing, Dirk and Ammoser, Hendrik and K{\"u}hnert, Christian},
  booktitle={Extreme events in nature and society},
  pages={319--348},
  year={2006},
  publisher={Springer}
}

@article{helbing2024co,
  title={Co-creating the future: participatory cities and digital governance},
  author={Helbing, Dirk and Mahajan, Sachit and Carpentras, Dino and Menendez, Monica and Pournaras, Evangelos and Thurner, Stefan and Verma, Trivik and Arcaute, Elsa and Batty, Michael and Bettencourt, Luis MA},
  journal={Philosophical Transactions A},
  volume={382},
  number={2285},
  pages={20240113},
  year={2024},
  publisher={The Royal Society}
}

@inproceedings{10.1145/1240624.1240704,
author = {Zimmerman, John and Forlizzi, Jodi and Evenson, Shelley},
title = {Research through design as a method for interaction design research in HCI},
year = {2007},
isbn = {9781595935939},
publisher = {Association for Computing Machinery},
address = {New York, NY, USA},
url = {https://doi.org/10.1145/1240624.1240704},
doi = {10.1145/1240624.1240704},
booktitle = {Proceedings of the SIGCHI Conference on Human Factors in Computing Systems},
pages = {493–502},
numpages = {10},
location = {San Jose, California, USA},
series = {CHI '07}
}

@inproceedings{buchenau2000experience,
  title={Experience prototyping},
  author={Buchenau, Marion and Suri, Jane Fulton},
  booktitle={Proceedings of the 3rd conference on Designing interactive systems: processes, practices, methods, and techniques},
  pages={424--433},
  year={2000}
}

@inproceedings{hartmann2006reflective,
  title={Reflective physical prototyping through integrated design, test, and analysis},
  author={Hartmann, Bj{\"o}rn and Klemmer, Scott R and Bernstein, Michael and Abdulla, Leith and Burr, Brandon and Robinson-Mosher, Avi and Gee, Jennifer},
  booktitle={Proceedings of the 19th annual ACM symposium on User interface software and technology},
  pages={299--308},
  year={2006}
}

@article{gulliksen2003key,
  title={Key principles for user-centred systems design},
  author={Gulliksen, Jan and G{\"o}ransson, Bengt and Boivie, Inger and Blomkvist, Stefan and Persson, Jenny and Cajander, {\AA}sa},
  journal={Behaviour and Information Technology},
  volume={22},
  number={6},
  pages={397--409},
  year={2003},
  publisher={Taylor \& Francis}
}

@inproceedings{vredenburg2002survey,
  title={A survey of user-centered design practice},
  author={Vredenburg, Karel and Mao, Ji-Ye and Smith, Paul W and Carey, Tom},
  booktitle={Proceedings of the SIGCHI conference on Human factors in computing systems},
  pages={471--478},
  year={2002}
}

@inproceedings{choudhuri2025guides,
  title={What Guides Our Choices? Modeling Developers' Trust and Behavioral Intentions Towards Genai},
  author={Choudhuri, Rudrajit and Trinkenreich, Bianca and Pandita, Rahul and Kalliamvakou, Eirini and Steinmacher, Igor and Gerosa, Marco and Sanchez, Christopher and Sarma, Anita},
  booktitle={2025 IEEE/ACM 47th International Conference on Software Engineering (ICSE)},
  pages={1691--1703},
  year={2025},
  organization={IEEE}
}

@article{brundage2020toward,
  title={Toward trustworthy AI development: mechanisms for supporting verifiable claims},
  author={Brundage, Miles and Avin, Shahar and Wang, Jasmine and Belfield, Haydn and Krueger, Gretchen and Hadfield, Gillian and Khlaaf, Heidy and Yang, Jingying and Toner, Helen and Fong, Ruth and others},
  journal={arXiv preprint arXiv:2004.07213},
  year={2020}
}

@inproceedings{wang2021autods,
  title={Autods: Towards human-centered automation of data science},
  author={Wang, Dakuo and Andres, Josh and Weisz, Justin D and Oduor, Erick and Dugan, Casey},
  booktitle={Proceedings of the 2021 CHI conference on human factors in computing systems},
  pages={1--12},
  year={2021}
}

@inproceedings{hase2020evaluating,
  title={Evaluating Explainable AI: Which Algorithmic Explanations Help Users Predict Model Behavior?},
  author={Hase, Peter and Bansal, Mohit},
  booktitle={Proceedings of the 58th Annual Meeting of the Association for Computational Linguistics},
  pages={5540--5552},
  year={2020}
}

@article{wright2021recast,
  title={RECAST: Enabling user recourse and interpretability of toxicity detection models with interactive visualization},
  author={Wright, Austin P and Shaikh, Omar and Park, Haekyu and Epperson, Will and Ahmed, Muhammed and Pinel, Stephane and Chau, Duen Horng and Yang, Diyi},
  journal={Proceedings of the ACM on human-computer interaction},
  volume={5},
  number={CSCW1},
  pages={1--26},
  year={2021},
  publisher={ACM New York, NY, USA}
}

@inproceedings{cheng2018human,
  title={Human behavior under emergency and its simulation modeling: a review},
  author={Cheng, Yixuan and Liu, Dahai and Chen, Jie and Namilae, Sirish and Thropp, Jennifer and Seong, Younho},
  booktitle={International conference on applied human factors and ergonomics},
  pages={313--325},
  year={2018},
  organization={Springer}
}

@article{kostakos2010brief,
  title={Brief encounters: Sensing, modeling and visualizing urban mobility and copresence networks},
  author={Kostakos, Vassilis and O'Neill, Eamonn and Penn, Alan and Roussos, George and Papadongonas, Dikaios},
  journal={ACM Transactions on Computer-Human Interaction (TOCHI)},
  volume={17},
  number={1},
  pages={1--38},
  year={2010},
  publisher={ACM New York, NY, USA}
}

@article{aguirre2011contributions,
  title={Contributions of social science to agent-based models of building evacuation},
  author={Aguirre, Benigno E and El-Tawil, Sherif and Best, Eric and Gill, Kimberly B and Fedorov, Vladimir},
  journal={Contemporary Social Science},
  volume={6},
  number={3},
  pages={415--432},
  year={2011},
  publisher={Taylor \& Francis}
}

@article{ligmann2014using,
  title={Using uncertainty and sensitivity analyses in socioecological agent-based models to improve their analytical performance and policy relevance},
  author={Ligmann-Zielinska, Arika and Kramer, Daniel B and Spence Cheruvelil, Kendra and Soranno, Patricia A},
  journal={PloS one},
  volume={9},
  number={10},
  pages={e109779},
  year={2014},
  publisher={Public Library of Science San Francisco, USA}
}

@article{schelling1971dynamic,
  title={Dynamic models of segregation},
  author={Schelling, Thomas C},
  journal={Journal of mathematical sociology},
  volume={1},
  number={2},
  pages={143--186},
  year={1971},
  publisher={Taylor \& Francis}
}

@article{an2012modeling,
  title={Modeling human decisions in coupled human and natural systems: Review of agent-based models},
  author={An, Li},
  journal={Ecological modelling},
  volume={229},
  pages={25--36},
  year={2012},
  publisher={Elsevier}
}

@inproceedings{macal2005tutorial,
  title={Tutorial on agent-based modeling and simulation},
  author={Macal, Charles M and North, Michael J},
  booktitle={Proceedings of the Winter Simulation Conference, 2005.},
  pages={14--pp},
  year={2005},
  organization={IEEE}
}

@article{joo2013agent,
  title={Agent-based simulation of affordance-based human behaviors in emergency evacuation},
  author={Joo, Jaekoo and Kim, Namhun and Wysk, Richard A and Rothrock, Ling and Son, Young-Jun and Oh, Yeong-gwang and Lee, Seungho},
  journal={Simulation Modelling Practice and Theory},
  volume={32},
  pages={99--115},
  year={2013},
  publisher={Elsevier}
}

@article{rozo2019modelling,
  title={Modelling building emergency evacuation plans considering the dynamic behaviour of pedestrians using agent-based simulation},
  author={Rozo, Kelly Rend{\'o}n and Arellana, Julian and Santander-Mercado, Alcides and Jubiz-Diaz, Maria},
  journal={Safety science},
  volume={113},
  pages={276--284},
  year={2019},
  publisher={Elsevier}
}

@article{chen2008agent,
  title={Agent-based modelling and simulation of urban evacuation: relative effectiveness of simultaneous and staged evacuation strategies},
  author={Chen, Xuwei and Zhan, F Benjamin},
  journal={Journal of the Operational Research Society},
  volume={59},
  number={1},
  pages={25--33},
  year={2008},
  publisher={Taylor \& Francis}
}

@article{helbing1995social,
  title={Social force model for pedestrian dynamics},
  author={Helbing, Dirk and Molnar, Peter},
  journal={Physical review E},
  volume={51},
  number={5},
  pages={4282},
  year={1995},
  publisher={APS}
}

@article{sun2016simple,
  title={Simple or complicated agent-based models? A complicated issue},
  author={Sun, Zhanli and Lorscheid, Iris and Millington, James D and Lauf, Steffen and Magliocca, Nicholas R and Groeneveld, J{\"u}rgen and Balbi, Stefano and Nolzen, Henning and M{\"u}ller, Birgit and Schulze, Jule and others},
  journal={Environmental Modelling \& Software},
  volume={86},
  pages={56--67},
  year={2016},
  publisher={Elsevier}
}

@article{deangelis2019decision,
  title={Decision-making in agent-based modeling: A current review and future prospectus},
  author={DeAngelis, Donald L and Diaz, Stephanie G},
  journal={Frontiers in Ecology and Evolution},
  volume={6},
  pages={237},
  year={2019},
  publisher={Frontiers Media SA}
}

@article{groeneveld2017theoretical,
  title={Theoretical foundations of human decision-making in agent-based land use models--A review},
  author={Groeneveld, J{\"u}rgen and M{\"u}ller, Birgit and Buchmann, Carsten M and Dressler, Gunnar and Guo, Cheng and Hase, Niklas and Hoffmann, Falk and John, Felix and Klassert, Christian and Lauf, Thomas and others},
  journal={Environmental modelling \& software},
  volume={87},
  pages={39--48},
  year={2017},
  publisher={Elsevier}
}

@article{gao2024large,
  title={Large language models empowered agent-based modeling and simulation: A survey and perspectives},
  author={Gao, Chen and Lan, Xiaochong and Li, Nian and Yuan, Yuan and Ding, Jingtao and Zhou, Zhilun and Xu, Fengli and Li, Yong},
  journal={Humanities and Social Sciences Communications},
  volume={11},
  number={1},
  pages={1--24},
  year={2024},
  publisher={Palgrave}
}

@article{gurcan2024llm,
  title={Llm-augmented agent-based modelling for social simulations: Challenges and opportunities},
  author={Gurcan, Onder},
  journal={arXiv preprint arXiv:2405.06700},
  year={2024}
}

@book{castillo2016big,
  title={Big crisis data: social media in disasters and time-critical situations},
  author={Castillo, Carlos},
  year={2016},
  publisher={Cambridge University Press}
}

@article{jia2021triadic,
  title={Triadic embeddedness structure in family networks predicts mobile communication response to a sudden natural disaster},
  author={Jia, Jayson S and Li, Yiwei and Lu, Xin and Ning, Yijian and Christakis, Nicholas A and Jia, Jianmin},
  journal={Nature communications},
  volume={12},
  number={1},
  pages={4286},
  year={2021},
  publisher={Nature Publishing Group UK London}
}

@article{shirado2020collective,
  title={Collective communication and behaviour in response to uncertain ‘Danger’in network experiments},
  author={Shirado, Hirokazu and Crawford, Forrest W and Christakis, Nicholas A},
  journal={Proceedings of the Royal Society A},
  volume={476},
  number={2237},
  pages={20190685},
  year={2020},
  publisher={The Royal Society Publishing}
}

@article{reuter2018fifteen,
  title={Fifteen years of social media in emergencies: a retrospective review and future directions for crisis informatics},
  author={Reuter, Christian and Kaufhold, Marc-Andr{\'e}},
  journal={Journal of contingencies and crisis management},
  volume={26},
  number={1},
  pages={41--57},
  year={2018},
  publisher={Wiley Online Library}
}

@article{thompson2017evacuation,
  title={Evacuation from natural disasters: a systematic review of the literature},
  author={Thompson, Rebecca R and Garfin, Dana Rose and Silver, Roxane Cohen},
  journal={Risk analysis},
  volume={37},
  number={4},
  pages={812--839},
  year={2017},
  publisher={Wiley Online Library}
}

@article{takayasu2015rumor,
  title={Rumor diffusion and convergence during the 3.11 earthquake: a Twitter case study},
  author={Takayasu, Misako and Sato, Kazuya and Sano, Yukie and Yamada, Kenta and Miura, Wataru and Takayasu, Hideki},
  journal={PLoS one},
  volume={10},
  number={4},
  pages={e0121443},
  year={2015},
  publisher={Public Library of Science San Francisco, CA USA}
}

@article{starbird2014rumors,
  title={Rumors, false flags, and digital vigilantes: Misinformation on twitter after the 2013 boston marathon bombing},
  author={Starbird, Kate and Maddock, Jim and Orand, Mania and Achterman, Peg and Mason, Robert M},
  journal={iConference 2014 proceedings},
  year={2014},
  publisher={iSchools}
}

@article{utz2013crisis,
  title={Crisis communication online: How medium, crisis type and emotions affected public reactions in the Fukushima Daiichi nuclear disaster},
  author={Utz, Sonja and Schultz, Friederike and Glocka, Sandra},
  journal={Public relations review},
  volume={39},
  number={1},
  pages={40--46},
  year={2013},
  publisher={Elsevier}
}

@book{drabek2012human,
  title={Human system responses to disaster: An inventory of sociological findings},
  author={Drabek, Thomas E},
  year={2012},
  publisher={Springer Science \& Business Media}
}

@article{moss2002policy,
  title={Policy analysis from first principles},
  author={Moss, Scott},
  journal={Proceedings of the National Academy of Sciences},
  volume={99},
  number={suppl\_3},
  pages={7267--7274},
  year={2002},
  publisher={National Academy of Sciences}
}

@book{morgan1990uncertainty,
  title={Uncertainty: a guide to dealing with uncertainty in quantitative risk and policy analysis},
  author={Morgan, Millett Granger and Henrion, Max and Small, Mitchell},
  year={1990},
  publisher={Cambridge university press}
}

@article{comfort2007crisis,
  title={Crisis management in hindsight: Cognition, communication, coordination, and control},
  author={Comfort, Louise K},
  journal={Public administration review},
  volume={67},
  pages={189--197},
  year={2007},
  publisher={Wiley Online Library}
}

@inproceedings{palen2010vision,
  title={A vision for technology-mediated support for public participation \& assistance in mass emergencies \& disasters},
  author={Palen, Leysia and Anderson, Kenneth M and Mark, Gloria and Martin, James and Sicker, Douglas and Palmer, Martha and Grunwald, Dirk},
  booktitle={ACM-BCS Visions of Computer Science 2010},
  year={2010},
  organization={BCS Learning \& Development}
}

@article{van2021evacuation,
  title={Evacuation behaviors and emergency communications: An analysis of real-world incident videos},
  author={van der Wal, C Natalie and Robinson, Mark A and de Bruin, W{\"a}ndi Bruine and Gwynne, Steven},
  journal={Safety science},
  volume={136},
  pages={105121},
  year={2021},
  publisher={Elsevier}
}

@article{nilsson2009social,
  title={Social influence during the initial phase of a fire evacuation—Analysis of evacuation experiments in a cinema theatre},
  author={Nilsson, Daniel and Johansson, Anders},
  journal={Fire safety journal},
  volume={44},
  number={1},
  pages={71--79},
  year={2009},
  publisher={Elsevier}
}

@article{templeton2023and,
  title={Who and what is trusted in fire incidents? The role of trust in guidance and guidance creators in resident response to fire incidents in high-rise residential buildings},
  author={Templeton, Anne and Nash, Claire and Spearpoint, Michael and Gwynne, Steve and Hui, Xie and Arnott, Matthew},
  journal={Safety science},
  volume={164},
  pages={106172},
  year={2023},
  publisher={Elsevier}
}

@article{templeton2024agent,
  title={Agent-based models of social behaviour and communication in evacuations: A systematic review},
  author={Templeton, Anne and Xie, Hui and Gwynne, Steve and Hunt, Aoife and Thompson, Pete and K{\"o}ster, Gerta},
  journal={Safety Science},
  volume={176},
  pages={106520},
  year={2024},
  publisher={Elsevier}
}

@article{waddell2004case,
  title={A case study in digital government: Developing and applying UrbanSim, a system for simulating urban land use, transportation, and environmental impacts},
  author={Waddell, Paul and Borning, Alan},
  journal={Social science computer review},
  volume={22},
  number={1},
  pages={37--51},
  year={2004},
  publisher={Sage Publications}
}

@article{lempert2002agent,
  title={Agent-based modeling as organizational and public policy simulators},
  author={Lempert, Robert},
  journal={Proceedings of the national academy of sciences},
  volume={99},
  number={suppl\_3},
  pages={7195--7196},
  year={2002},
  publisher={National Academy of Sciences}
}

\appendix

\section{Appendix}
\setcounter{table}{0}
\setcounter{figure}{0}
\renewcommand{\thetable}{A\arabic{table}}
\renewcommand{\thefigure}{A\arabic{figure}}

\subsection{Co-creating Stakeholder Maps and Process Maps}
\begin{figure}[H]
  \centering 
  \includegraphics[width=0.5\textwidth]{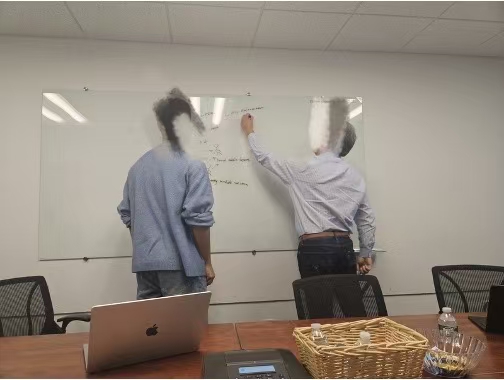}
  \caption{Developers and policymakers \textit{P1} collaboratively co-creating a stakeholder map and process map during a policy meeting. The session supported knowledge elicitation of roles, responsibilities, and workflows in emergency preparedness.}
  \Description{A photo of two people seated at a table in a meeting room, working together on a white board. They are sketching maps with markers to document stakeholders and processes for emergency preparedness.}
  \label{fig:co_creation}
\end{figure}

\begin{figure}[H]
  \centering 
  \includegraphics[width=0.9\textwidth]{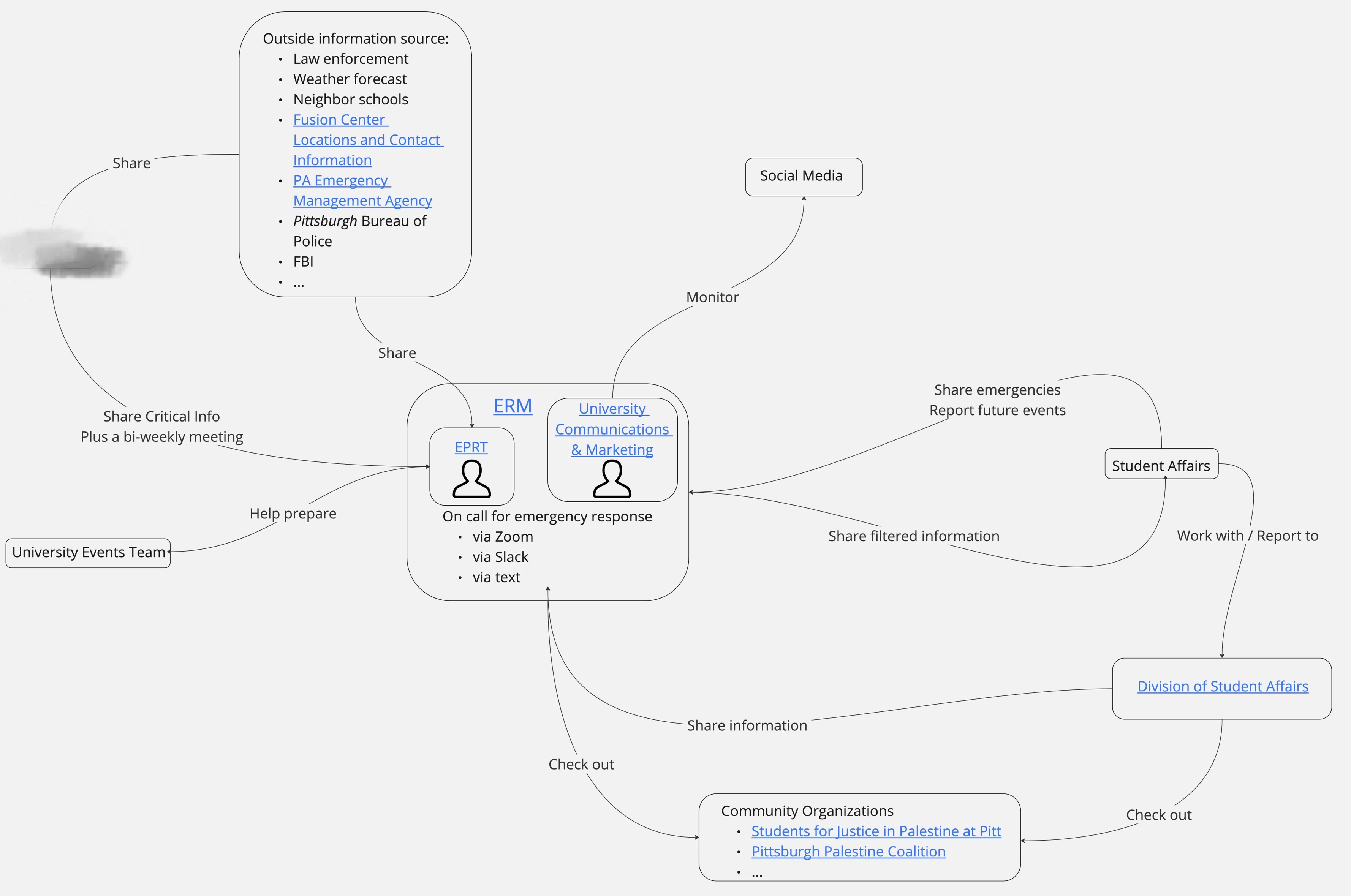}
  \caption{Stakeholder map of emergency preparedness and response at the university, co-created with policymakers to identify actors, roles, and relationships shaping decision-making and information flows during crises.}
  \Description{A network diagram showing university units, local government offices, emergency responders, and student groups, with labeled connections illustrating communication and authority relationships in emergency preparedness.}
  \label{fig:stakeholder_map}
\end{figure}

\begin{figure}[H]
  \centering 
  \includegraphics[width=0.9\textwidth]{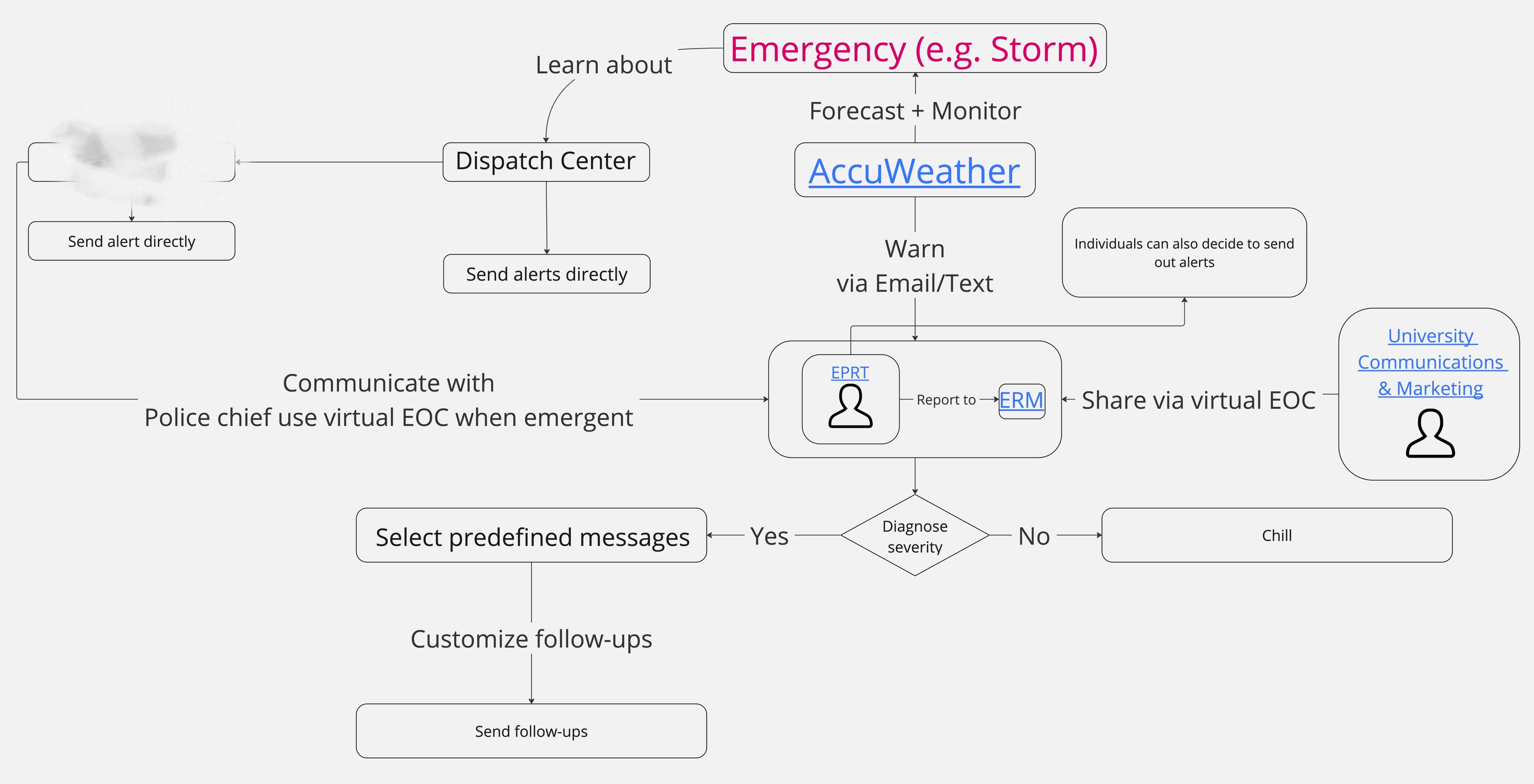}
  \caption{Process diagram of emergency preparedness and response at the university, documenting sequential phases from preparedness planning to active response and recovery, and highlighting interdependencies among roles.}
  \Description{A flowchart with boxes and arrows showing stages of preparedness (planning, drills), active response (alerts, evacuation), and recovery (debrief, after-action reporting), with actors assigned at each stage.}
  \label{fig:process_map_1}
\end{figure}

\begin{figure}[H]
  \centering 
  \includegraphics[width=0.9\textwidth]{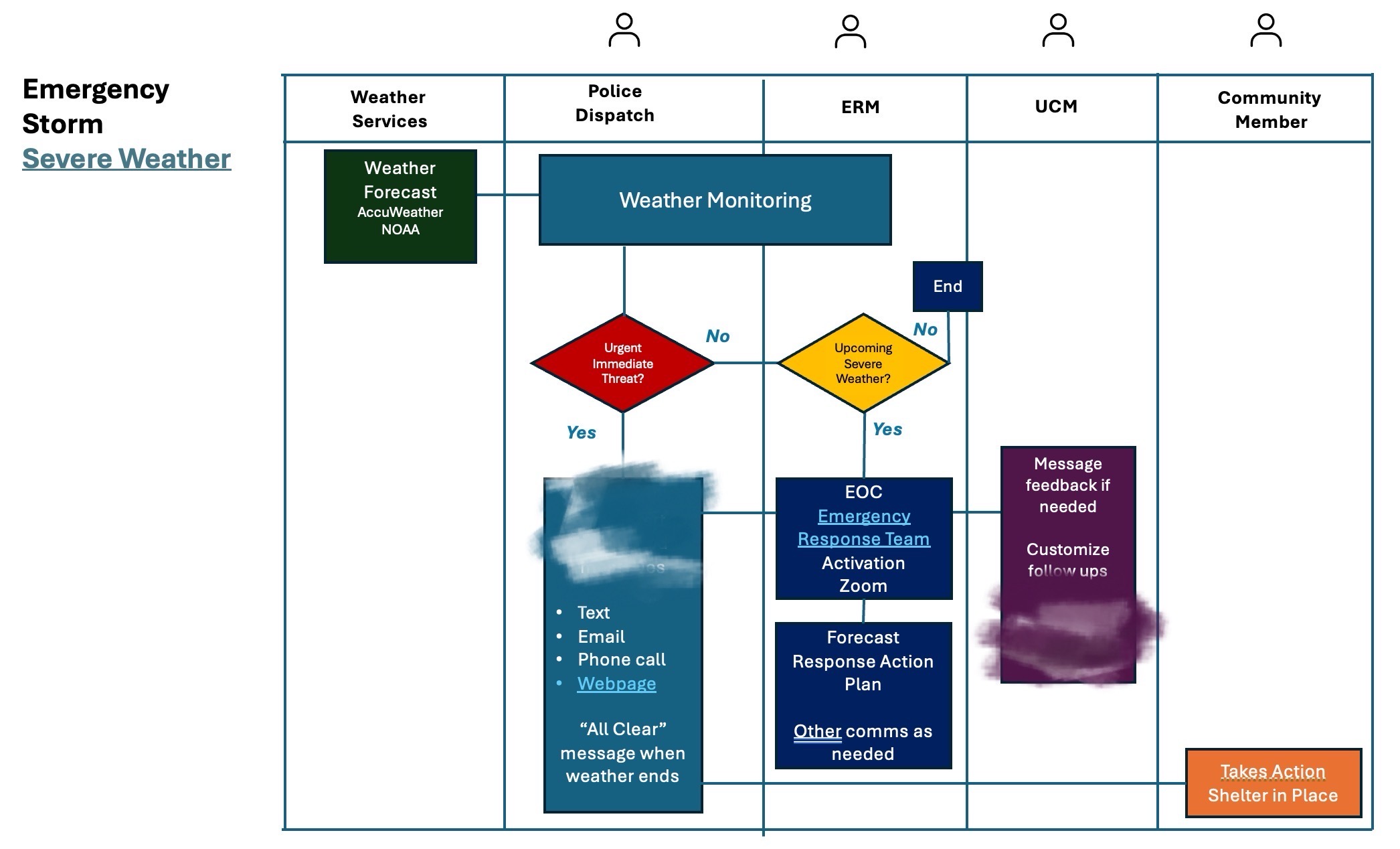}
  \caption{Process diagram of emergency preparedness and response tailored for severe weather scenarios, illustrating protocols for monitoring, issuing warnings, coordinating shelter, and resuming normal operations.}
  \Description{A process map with steps beginning at weather monitoring, progressing through warning issuance, shelter coordination, and eventual campus reopening, with designated responsibilities for staff and emergency services.}
  \label{fig:process_map_2}
\end{figure}

\begin{figure}[H]
  \centering 
  \includegraphics[width=0.9\textwidth]{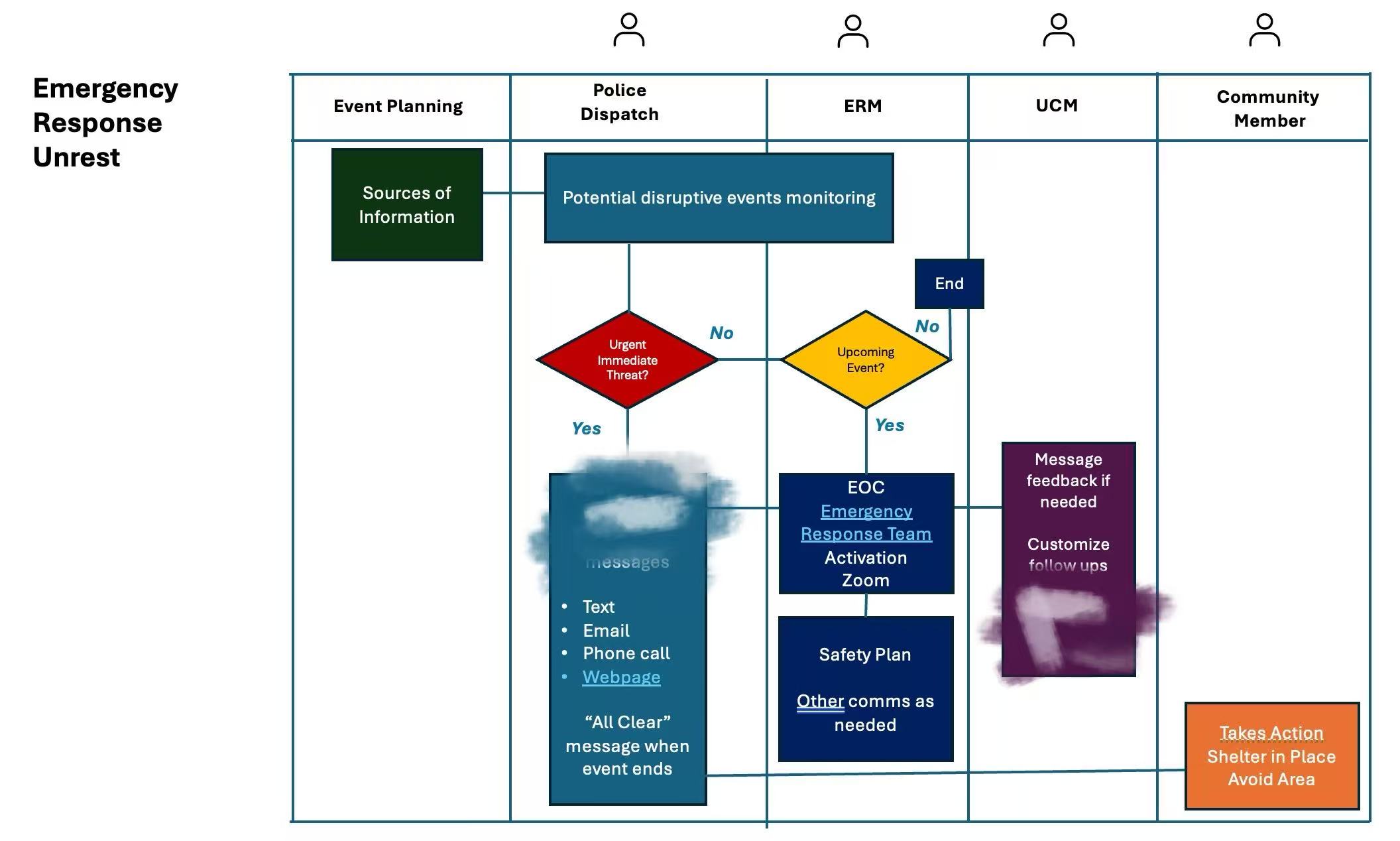}
  \caption{Process diagram of emergency preparedness and response tailored for unrest or bomb threat scenarios, showing decision pathways for lockdowns, communication flows, and coordination with law enforcement.}
  \Description{A flowchart with branching paths showing actions under unrest or bomb threat: initiating lockdowns, coordinating with police, informing students and staff, and transitioning from response to resolution.}
  \label{fig:process_map_3}
\end{figure}

\subsection{System Description: Iteration 1} \label{appendix:system_description_iteration_1}
\subsubsection{Misinterpretation Simulation System}
The simulation system implements a multi-stage pipeline designed to evaluate how university announcements may be misinterpreted by students with diverse cognitive and behavioral characteristics. The system operates through four primary phases: agent generation, message creation, interpretation simulation, and misinterpretation assessment.

\textbf{Agent generation and characterization.}
The system begins by generating a diverse population of simulated students, each represented as an agent with distinct personal characteristics. Each agent is assigned a randomized name and a detailed persona that encompasses various psychological and behavioral traits. The persona specifications explicitly include the individual's propensity for misinterpreting information, ranging from highly analytical students who carefully parse communications to individuals who may jump to conclusions or misread contextual cues. Additional persona elements may include academic background, stress levels, prior experiences with institutional communications, attention to detail, and emotional reactivity patterns.

\textbf{Message corpus development.}
In parallel to agent creation, the system generates a corpus of representative university announcement messages. These messages are designed to reflect typical institutional communications that students might receive, spanning various topics such as policy changes, event notifications, academic deadlines, or administrative updates.

\textbf{Interpretation and reaction simulation.}
The core simulation phase involves exposing each generated agent to each message in the corpus. For every agent-message pair, the system simulates the student's cognitive processing by generating two outputs: an interpretation of the message based on the agent's persona, and a behavioral reaction stemming from that interpretation. The interpretation represents the agent's internal understanding of the message content, filtered through their individual characteristics and biases. The reaction component captures how the student would likely respond to the message given their specific interpretation and personality traits.

\textbf{Misinterpretation scoring and assessment.}
Following the interpretation simulation, the system employs an independent assessment mechanism to evaluate the degree of misinterpretation present in each agent's response. This assessment compares the agent's interpretation and reaction against the original message content to assign a numerical misinterpretation score. The scoring system operates on a scale from zero to one hundred, where zero indicates perfect comprehension and one hundred represents complete misunderstanding of the intended message.

\textbf{Extreme response generation.}
For cases where the misinterpretation score exceeds a threshold value, indicating significant misunderstanding, the system generates an additional extreme reaction scenario. This component simulates how the student might respond if their misinterpretation led to heightened emotional or behavioral responses, providing insight into potential worst-case scenarios for communication failures.

\begin{algorithm}
\caption{Pseudo Script: Misinterpretation Simulation System}
\begin{algorithmic}[1]
\Require Number of agents $N$, Number of messages $M$
\Ensure Misinterpretation analysis results for all agent-message pairs

\State \textbf{Phase 1: Agent and Message Generation}
\State $A \gets \{\}$ \Comment{Initialize agent set}
\For{$i = 1 \to N$}
    \State $name_i \gets$ GenerateRandomName()
    \State $persona_i \gets$ GeneratePersona(misinterpretation\_tendency)
    \State $A \gets A \cup \{(name_i, persona_i)\}$
\EndFor

\State $Messages \gets$ GenerateMessageCorpus($M$) \Comment{Generate university announcements}

\State \textbf{Phase 2: Interpretation Simulation}
\State $Results \gets \{\}$ \Comment{Initialize results storage}
\ForAll{$message_j \in Messages$}
    \ForAll{$agent_i \in A$}
        \State $interpretation_{i,j} \gets$ SimulateInterpretation($agent_i$, $message_j$)
        \State $reaction_{i,j} \gets$ SimulateReaction($agent_i$, $message_j$, $interpretation_{i,j}$)
        
        \State \textbf{Phase 3: Misinterpretation Assessment}
        \State $score_{i,j} \gets$ AssessMisinterpretation($message_j$, $interpretation_{i,j}$, $reaction_{i,j}$)
        
        \State \textbf{Phase 4: Extreme Response Generation}
        \If{$score_{i,j} > threshold$}
            \State $extreme_{i,j} \gets$ GenerateExtremeReaction($agent_i$, $message_j$, $interpretation_{i,j}$)
        \Else
            \State $extreme_{i,j} \gets \emptyset$
        \EndIf
        
        \State $Results[j] \gets Results[j] \cup \{(agent_i, interpretation_{i,j}, reaction_{i,j}, score_{i,j}, extreme_{i,j})\}$
    \EndFor
\EndFor

\Return $Results$ \Comment{Complete misinterpretation analysis dataset}
\end{algorithmic}
\end{algorithm}

\subsubsection{Propagation Simulation System}
The LLM agent simulation framework implements a multi-agent system designed to model information propagation and collective decision-making behaviors in response to security-related scenarios. The system consists of three core components: autonomous agents with individual personas, information diffusion mechanisms, and environmental orchestration.

\textbf{Agent architecture and decision-making.} Each agent possesses a unique identity with a name and detailed persona that influences their decision-making patterns. Agents operate through a structured three-phase decision pipeline when processing information. In the decision phase, agents evaluate all received information using their persona and accumulated knowledge to determine their reaction from three possible states: idle behavior, active information spreading, or evacuation from the simulation. When agents choose to spread information, they enter an action phase where they generate original content based on their understanding of the situation and their personal characteristics. Finally, agents undergo an update phase where they incorporate feedback about simulation outcomes into their knowledge base, enabling learning and adaptation for future scenarios.

The agent decision-making process utilizes large language models with structured prompts tailored to each agent's persona. Agents maintain separate conversation histories for each decision type, ensuring contextual consistency while preventing interference between different cognitive processes. The reward mechanism allows agents to accumulate insights from environmental feedback, creating memory effects where past experiences influence future behavior patterns. Agents can transition between active and inactive states, with evacuation representing a permanent withdrawal from the simulation.

\textbf{Information diffusion mechanisms.} The system implements two distinct diffusion strategies to study the effects of content moderation on information propagation. The moderated diffusion approach intercepts all agent-generated content and applies standardized warning labels indicating potential misinformation before distribution. Specifically, content is prefixed with explicit warnings stating ``WARNING: This piece of content might contain misinformation'' followed by the original agent message. The unmoderated diffusion approach transmits agent-generated content without any modification, serving as a control condition.

Both diffusion mechanisms employ identical probabilistic distribution patterns. When an agent selects the spreading action, the diffusion system randomly samples 70\% of all agents in the simulation as recipients for the generated content. This sampling approach creates realistic information propagation patterns where content does not reach the entire population simultaneously. The 70\% coverage rate ensures substantial information spread while maintaining realistic constraints on individual agent reach within social networks.

\textbf{Environmental coordination and scenario management.} The environmental system orchestrates complex multi-round simulations through structured scenario scripts that define information injection patterns, timing, and feedback mechanisms. These scripts specify which agents receive initial information, the source attribution for credibility modeling (such as news organizations or individual sources), and the specific content payloads representing rumors, official announcements, or other information types.

Simulation execution proceeds through discrete rounds where the environment first distributes scripted information to designated agent subsets. Active agents then process their inputs and make decisions according to their decision pipeline. Any agents choosing to spread information generate content that enters the diffusion mechanism, creating secondary information waves that become inputs for subsequent rounds. This process continues until all scripted rounds complete, generating cascading information effects throughout the agent population.

The system implements section-based feedback where agents receive environmental outcomes based on their final states at the end of each scenario section. Agents learn whether their decisions were appropriate given the actual nature of the security threat, enabling behavioral adaptation across multiple scenario sections.

\textbf{Information propagation dynamics and network effects.} Information flow within the system emerges organically from agent decisions rather than following predefined network topologies. The system creates dynamic communication networks where information pathways form based on which agents choose to spread content and which agents are randomly selected as recipients. This approach models realistic social media environments where information sharing creates temporary communication channels between individuals.

The propagation mechanism generates compound effects where initial information triggers multiple rounds of agent responses, decisions, and subsequent information generation. Agent-generated content influences recipient decision-making, potentially creating viral information cascades where rumors or alerts spread exponentially through the population. The random sampling approach ensures that different agents may receive different combinations of information, leading to diverse response patterns and realistic heterogeneity in population-level outcomes.

\begin{algorithm}
\caption{Pseudo Script: Propagation Simulation System}
\label{alg:social_simulation}
\begin{algorithmic}[1]
\Require $A = \{a_1, a_2, \ldots, a_n\}$ \Comment{Set of agents with personas}
\Require $S$ \Comment{Scenario script with sections and rounds}
\Require $D$ \Comment{Diffusion mechanism (moderated or unmoderated)}

\State Initialize all agents $a_i$ with personas, decision histories, and reward sets
\State Set all agent states to ACTIVE

\For{each section $s$ in $S$}
    \State $propagated\_info \gets \emptyset$
    
    \For{each round $r$ in $s$}
        \State $current\_round\_actions \gets \emptyset$
        \State $round\_inputs \gets$ Combine($r$, $propagated\_info$) \Comment{Merge scripted and propagated info}
        
        \For{each agent $a_i \in A$ where $state(a_i) = \text{ACTIVE}$ and $a_i$ has inputs}
            \State $inputs_i \gets round\_inputs[a_i]$
            \State $context \gets$ BuildContext($a_i.persona$, $a_i.rewards$, $inputs_i$)
            \State $decision_i \gets$ LLMCall($a_i.decision\_history$, $context$)
            
            \If{$decision_i = \text{SPREAD}$}
                \State $content_i \gets$ LLMCall($a_i.action\_history$, $context$)
                \State Add $\{role: a_i.name, content: content_i\}$ to $current\_round\_actions$
            \ElsIf{$decision_i = \text{EVACUATE}$}
                \State $state(a_i) \gets \text{INACTIVE}$
            \EndIf
        \EndFor
        
        \State $propagated\_info \gets \emptyset$
        \For{each action $act$ in $current\_round\_actions$}
            \State $recipients \gets$ RandomSample($A$, $0.7 \times |A|$) \Comment{70\% coverage}
            \If{$D$ is moderated}
                \State $content \gets$ "WARNING: misinformation. " + $act.content$
            \Else
                \State $content \gets act.content$
            \EndIf
            \State Add $\{recipients: recipients, source: act.role, content: content\}$ to $propagated\_info$
        \EndFor
    \EndFor
    
    \If{$s$ is not the last section}
        \For{each agent $a_i \in A$}
            \State $result_i \gets$ GetSectionResult($s$, $state(a_i)$)
            \State $knowledge \gets$ LLMCall($a_i.update\_history$, BuildUpdateContext($result_i$))
            \State $a_i.rewards \gets knowledge$
            \State Clear agent histories and reset $state(a_i) \gets \text{ACTIVE}$
        \EndFor
    \EndIf
\EndFor
\end{algorithmic}
\end{algorithm}

\subsection{System Description: Final Iteration} \label{appendix:system_description_final_iteration}

\subsubsection{Agent Population Generation}

The simulation employs a \textbf{two-stage persona generation process} to create realistic and diverse agent populations representing commencement attendees. The first stage generates individual student personas through parallel API calls to language models, while the second stage creates accompanying family and friend personas for students requiring social connections.

Student persona generation operates across ten academic disciplines with predetermined enrollment distributions, totaling 2,928 base student personas. Each generation request includes the academic major as context to produce discipline-appropriate background profiles. The system generates an additional 44 students with accessibility requirements, distributed proportionally across all academic programs.

The persona generation algorithm assigns unique demographic characteristics including names, detailed background descriptions encompassing academic interests and personal circumstances, accessibility requirements, and group affiliation patterns. Students with accessibility needs receive explicit mobility considerations that influence their movement parameters during simulation execution.

The second generation stage creates social network structures through a stratified assignment process. The system randomly partitions the 2,928 students into three categories: 2,000 students who attend with family members or friends (requiring additional persona generation), 800 students who form friend groups with other students, and 128 students who attend individually. For the family/friend category, the system generates between 1-8 additional personas per student using targeted prompts that specify relationship types and demographic coherence with the primary student.

Friend group formation employs stochastic clustering with group sizes ranging from 3-10 members, randomly sampling from the designated student pool until all members receive group assignments. Each generated group establishes bidirectional membership lists that define communication channels and decision-making units during simulation execution.

The complete population generation process produces approximately 13,000 total agents with explicit social network topologies, realistic demographic distributions, and heterogeneous accessibility requirements that directly influence simulation behavior patterns.

\subsubsection{Agent Architecture and Initialization}

The simulation instantiates individual agents representing commencement attendees, each configured with distinct behavioral parameters and social network affiliations. Each agent maintains a comprehensive profile including demographic information, academic major, accessibility requirements, and explicit group membership identifiers that define their social connections within the simulation.

The system categorizes agents into four distinct behavioral classes that determine their decision-making protocols: students with family and friends outside the stadium, students with friends inside the venue, students attending alone, and family members or friends of graduates. These classifications directly influence the agent's response patterns and group interaction capabilities.

Agent initialization occurs at predetermined coordinates within a 2400×1200 pixel discrete coordinate system representing the stadium layout. The coordinate space encompasses geometrically defined regions including eight numbered seating sections arranged in a 2×4 grid, interconnecting pathways with specified widths, designated family and accessibility areas positioned along the stadium perimeter, a central rectangular stage obstacle, and four exit points located at coordinates (20,20), (2380,20), (20,1180), and (2380,600).

Each agent operates with a 20-pixel visibility radius for environmental detection and maintains internal state variables tracking their current position, movement target coordinates, decision history, and group chat message logs. The system assigns each agent an accessibility flag that modifies their movement parameters and a visibility radius parameter that determines their environmental awareness range.

\begin{algorithm}
\caption{Commencement Emergency Response Simulation System (Part 1: Initialization \& Population)}
\begin{algorithmic}[1]

\Procedure{InitializeSimulation}{}
    \State $\textit{canvas} \leftarrow$ Create2DSpace(2400, 1200)
    \State $\textit{stadium\_features} \leftarrow$ DefineStadiumLayout()
    \State $\textit{coordinators} \leftarrow$ PlaceCoordinators(50, $\textit{canvas}$)
    \State $\textit{agents} \leftarrow$ GenerateAgentPopulation()
    \State $\textit{round} \leftarrow 0$
    \State \Return $\textit{agents}, \textit{canvas}, \textit{coordinators}$
\EndProcedure

\Procedure{GenerateAgentPopulation}{}
    \State $\textit{students} \leftarrow \emptyset$
    \For{$\textit{major}$ in $\{$Engineering(720), Business(240), $\ldots\}$}
        \State $\textit{personas} \leftarrow$ GeneratePersonasAsync($\textit{major}$)
        \State $\textit{students} \leftarrow \textit{students} \cup \textit{personas}$
    \EndFor
    \State $\textit{population} \leftarrow$ CreateSocialNetworks($\textit{students}$)
    \State \Return $\textit{population}$
\EndProcedure

\Procedure{CreateSocialNetworks}{$\textit{students}$}
    \State $\textit{others\_list}, \textit{friends\_list}, \textit{alone\_list} \leftarrow$ RandomPartition($\textit{students}$, [2000, 800, 128])
    \For{$\textit{student}$ in $\textit{others\_list}$}
        \State $\textit{family\_count} \leftarrow$ Random(1, 8)
        \State $\textit{family} \leftarrow$ GenerateFamilyPersonas($\textit{student}$, $\textit{family\_count}$)
        \State SetGroupMembership($\textit{student} \cup \textit{family}$)
    \EndFor
    \For{$\textit{remaining}$ in $\textit{friends\_list}$}
        \State $\textit{group\_size} \leftarrow$ Random(3, 10)
        \State $\textit{group} \leftarrow$ Sample($\textit{remaining}$, $\textit{group\_size}$)
        \State SetGroupMembership($\textit{group}$)
    \EndFor
    \State \Return $\textit{others\_list} \cup \textit{friends\_list} \cup \textit{alone\_list}$
\EndProcedure

\end{algorithmic}
\end{algorithm}

\subsubsection{Decision-Making Process and Response Formats}

The simulation implements two distinct decision-making protocols differentiated by agent classification and social context. Non-student-alone agents participate in structured group discussion rounds using a multi-field response format that captures both individual decision state and social communication. These agents generate responses containing a boolean decision indicator, an optional destination selection from a predefined enumeration of twelve possible locations, and a natural language message for intra-group communication.

The destination enumeration includes four exit locations (Exit 1 through Exit 4), five stadium region descriptions (North/South/West/East track areas, South bleachers area), and three social gathering areas (West/East family and friends areas, West seating sections area). Each destination maps to specific coordinate generation algorithms that produce either fixed exit coordinates or randomized positions within defined regional boundaries.

Student-alone agents bypass group consultation mechanisms and utilize a simplified response protocol that produces only a destination selection without accompanying social messaging. This streamlined process eliminates group coordination overhead while maintaining equivalent environmental input processing.

Both decision protocols receive identical contextual information packages generated through environmental analysis algorithms. These packages contain structured descriptions of the agent's current stadium location, enumerated nearby physical features within the visibility radius, directional and distance information for all visible agents, coordinator proximity notifications with specific exit recommendations, and ranked distance calculations to all four exits with cardinal direction indicators.

\subsubsection{Environmental Context Generation}

The simulation constructs detailed environmental descriptions through systematic analysis of agent position relative to stadium features and other attendees. The context generation algorithm first identifies the agent's current location by testing point-in-rectangle containment against all defined stadium features, producing location-specific descriptions that vary based on feature type.

For agents positioned within seating sections, the system calculates row and column positions using coordinate offset arithmetic and unit spacing parameters. Agents on pathways receive directional information about pathway endpoints and connecting features. The algorithm generates specialized descriptions for agents in family areas, accessibility zones, or open spaces between major features.

Proximity detection operates through distance calculations between agent positions and all stadium features, filtering results by the agent's visibility radius. The system computes cardinal directions using arctangent calculations and maps angular ranges to eight-direction compass bearings. Distance measurements undergo categorical classification into descriptive ranges: extremely close (less than 50 pixels), near (50-150 pixels), moderately far (150-400 pixels), far (400-800 pixels), and very far (greater than 800 pixels).

The system performs comprehensive exit ranking by calculating Euclidean distances from the agent's position to all four exit coordinates, sorting results by proximity, and generating formatted descriptions that include both distance categories and directional bearings. This exit information provides agents with consistent spatial orientation data for navigation decision-making.

\begin{algorithm}
\caption{Commencement Emergency Response Simulation System (Part 2a: Main Loop Core)}
\begin{algorithmic}[1]

\Procedure{SimulationMainLoop}{$\textit{agents}$, $\textit{canvas}$, $\textit{coordinators}$}
    \While{$\exists \textit{agent} \in \textit{agents} : \textit{agent.state} \neq \text{EXITED}$}
        \State $\textit{round} \leftarrow \textit{round} + 1$
        
        \Comment{Phase 1: State Transition Analysis}
        \State $\textit{resumed\_agents} \leftarrow$ CheckGroupArrivals($\textit{agents}$)
        \State $\textit{influenced\_groups} \leftarrow$ CheckCoordinatorInfluence($\textit{agents}$, $\textit{coordinators}$)
        \State ResetInfluencedGroups($\textit{influenced\_groups}$)
        
        \Comment{Phase 2: Decision Processing}
        \State $\textit{discussing\_agents} \leftarrow \{a \in \textit{agents} : a.\text{state} = \text{DISCUSSING}\}$
        \State $\textit{decisions} \leftarrow$ ProcessDecisionsConcurrently($\textit{discussing\_agents}$)
        \State UpdateAgentStates($\textit{decisions}$)
        
        \Comment{Phase 3: Movement Processing}
        \State $\textit{moving\_agents} \leftarrow \{a \in \textit{agents} : a.\text{state} = \text{MOVING}\}$
        \State $\textit{density\_map} \leftarrow$ CalculateDensity($\textit{moving\_agents}$)
        \State ExecuteMovement($\textit{moving\_agents}$, $\textit{density\_map}$)
        \State CheckDestinationArrivals($\textit{moving\_agents}$)
        
        \Comment{Phase 4: Data Logging}
        \State LogRoundData($\textit{round}$, $\textit{agents}$, $\textit{decisions}$)
    \EndWhile
\EndProcedure

\end{algorithmic}
\end{algorithm}

\subsubsection{Movement Mechanics and Pathfinding}

Agent movement operates through a coordinate-based pathfinding system that translates abstract destination selections into specific target coordinates and executes movement through obstacle-aware navigation algorithms. The destination resolution process maps each of the twelve possible destination types to coordinate generation functions that produce either deterministic exit positions or stochastic coordinates within defined regional boundaries.

Regional destinations employ bounded random coordinate generation with feature-aware constraints. For example, "North side of the stadium, track area" generates coordinates within the northern boundary rectangle while explicitly avoiding intersection with the central stage obstacle through point-in-rectangle collision detection.

The movement calculation system processes agent motion through a multi-stage algorithm that accounts for movement constraints, obstacle avoidance, and environmental factors. Base movement speed varies according to agent characteristics: standard agents move at 24 pixels per round, accessibility-flagged agents move at 16 pixels per round, and all agents receive speed modifications based on environmental conditions.

Density-based speed adjustment operates through a graduated reduction system that counts nearby agents within a configurable radius and applies speed penalties. The algorithm maintains maximum speed when fewer than 4 agents occupy the proximity zone, linearly reduces speed as nearby agent count increases toward 30 agents, and enforces minimum speed thresholds to prevent complete movement cessation in crowded conditions.

Coordinator proximity provides speed enhancement, increasing movement rates to 32 pixels per round for standard agents and 16 pixels per round for accessibility agents when within 50 pixels of any coordinator entity. This mechanism simulates urgency responses to authority figure instructions.

The pathfinding algorithm implements vector-based movement with obstacle sliding behavior. The system calculates unit vectors toward target coordinates, detects intersection points with obstacle boundaries using line-segment-rectangle intersection calculations, and applies sliding movement along obstacle edges when direct paths are blocked. Special handling prevents agents from entering the stage area unless they begin their movement within that region.

\subsubsection{Coordinator Influence and Behavioral Modification}

The simulation deploys 50 coordinator entities at predetermined strategic positions throughout the stadium layout, each configured with specific exit recommendation parameters. Coordinator placement follows a spatial distribution pattern: northern coordinators recommend Exit 1, southern coordinators suggest Exit 3, with positioning designed to provide coverage across major circulation areas including pathway intersections, seating section perimeters, and family gathering zones.

The coordinator influence mechanism operates through proximity detection coupled with destination conflict analysis. When agents enter the 50-pixel influence radius around any coordinator, the system compares the agent's current destination against the coordinator's programmed exit recommendation. Destination mismatches trigger the influence protocol, which applies probabilistic behavioral modification based on a configurable reaction probability parameter.

Upon influence activation, the system implements group-level behavioral reset mechanisms. For student-alone agents, the influence directly affects the individual agent. For group-affiliated agents, the influence cascades to all members of the agent's social group, regardless of their individual proximity to coordinators. This collective influence model simulates realistic group decision-making dynamics where authority recommendations affect entire social units.

The behavioral reset process transitions all affected agents from their current movement or waiting states back to discussion state, clears their existing destination selections and movement targets, and injects the coordinator's exit recommendation into their next decision-making round as environmental context. The system tracks these influence events in the simulation log with detailed records of affected agents, original destinations, coordinator suggestions, and agent positions at the time of influence.

\begin{algorithm}
\caption{Commencement Emergency Response Simulation System (Part 2b: Coordinator Influence \& Decisions)}
\begin{algorithmic}[1]

\Procedure{CheckCoordinatorInfluence}{$\textit{agents}$, $\textit{coordinators}$}
    \State $\textit{influenced\_groups} \leftarrow \emptyset$
    \For{$\textit{agent}$ in $\{a \in \textit{agents} : a.\text{state} = \text{MOVING}\}$}
        \For{$\textit{coord}$ in $\textit{coordinators}$}
            \If{Distance($\textit{agent.position}$, $\textit{coord.position}$) $\leq 50$}
                \If{$\textit{agent.destination} \neq \textit{coord.suggested\_exit}$}
                    \If{Random() $< 0.5$}
                        \State $\textit{influenced\_groups} \leftarrow \textit{influenced\_groups} \cup \{\textit{agent.group}\}$
                    \EndIf
                \EndIf
            \EndIf
        \EndFor
    \EndFor
    \State \Return $\textit{influenced\_groups}$
\EndProcedure

\Procedure{ProcessDecisionsConcurrently}{$\textit{discussing\_agents}$}
    \State $\textit{tasks} \leftarrow \emptyset$
    \For{$\textit{agent}$ in $\textit{discussing\_agents}$}
        \State $\textit{context} \leftarrow$ GenerateEnvironmentalContext($\textit{agent}$)
        \If{$\textit{agent.type} = \text{student\_alone}$}
            \State $\textit{task} \leftarrow$ CreateDecisionTask($\textit{agent}$, $\textit{context}$)
        \Else
            \State $\textit{group\_messages} \leftarrow$ GetGroupChatHistory($\textit{agent.group}$)
            \State $\textit{task} \leftarrow$ CreateDiscussionTask($\textit{agent}$, $\textit{context}$, $\textit{group\_messages}$)
        \EndIf
        \State $\textit{tasks} \leftarrow \textit{tasks} \cup \{\textit{task}\}$
    \EndFor
    \State $\textit{responses} \leftarrow$ ExecuteTasksConcurrently($\textit{tasks}$, max\_concurrent=2000)
    \State \Return $\textit{responses}$
\EndProcedure

\Procedure{GenerateEnvironmentalContext}{$\textit{agent}$}
    \State $\textit{current\_location} \leftarrow$ IdentifyStadiumFeature($\textit{agent.position}$)
    \State $\textit{nearby\_features} \leftarrow$ FindFeaturesInRadius($\textit{agent.position}$, 20)
    \State $\textit{nearby\_agents} \leftarrow$ FindAgentsInRadius($\textit{agent.position}$, 20)
    \State $\textit{exit\_rankings} \leftarrow$ RankExitsByDistance($\textit{agent.position}$)
    \State $\textit{coordinator\_info} \leftarrow$ CheckCoordinatorProximity($\textit{agent.position}$)
    \State $\textit{description} \leftarrow$ GenerateNaturalLanguageDescription(
        \Statex \hspace{4em} $\textit{current\_location}$, $\textit{nearby\_features}$, $\textit{nearby\_agents}$, 
        \Statex \hspace{4em} $\textit{exit\_rankings}$, $\textit{coordinator\_info}$)
    \State \Return $\textit{description}$
\EndProcedure

\end{algorithmic}
\end{algorithm}

\subsubsection{State Management and Group Coordination}

Agent state management operates through a four-state finite state machine with explicit transition conditions and group synchronization mechanisms. The states comprise: discussing (participating in decision-making rounds), moving (navigating toward selected destinations), waiting (paused at intermediate destinations), and exited (reached final exit and removed from active simulation).

The group coordination system maintains data structures tracking destination selections and arrival status for each social group, identified by sorted member ID tuples. When agents select destinations, the system records both the choice and the selecting agent in group-specific tracking tables. As agents reach their target coordinates (determined by Euclidean distance calculations with a 50-pixel tolerance threshold), the system updates arrival status records.

Automatic discussion resumption occurs when all group members who selected identical intermediate destinations successfully arrive at those locations. The coordination algorithm periodically scans all tracked groups, identifies destinations where the set of selecting agents equals the set of arrived agents, and transitions all relevant agents from waiting state back to discussing state. This mechanism ensures group cohesion during multi-stage navigation without requiring explicit coordination messages.

Special handling accommodates student-alone agents who bypass group coordination requirements. These agents automatically transition from waiting to discussing states without requiring group member synchronization, preventing indefinite waiting conditions for agents without social dependencies.

\subsubsection{Round-Based Execution and Concurrency Management}

The simulation operates through discrete \textbf{round-based execution cycles} that process agent decisions and movements in structured phases. Each round begins with state transition analysis, identifying agents eligible for discussion resumption based on group arrival status and processing coordinator influence effects on currently moving agents.

The discussion processing phase handles all agents in discussing state through concurrent API request management. The system implements semaphore-based concurrency control with configurable limits (defaulting to 2000 concurrent requests) and rate limiting mechanisms to manage external language model API interactions. Discussion tasks execute asynchronously with timeout handling and retry logic for failed requests.

Response processing extracts structured decision data from language model outputs, updates agent internal states based on destination selections, and maintains group chat message histories for subsequent rounds. The system validates destination selections against the predefined enumeration and applies fuzzy string matching to handle response variations.

Movement processing operates on all agents in moving state simultaneously, calculating density-based speed adjustments through spatial proximity analysis and updating agent coordinates via pathfinding algorithms. The system performs destination arrival checking after position updates and transitions agents to appropriate successor states based on destination types (exits trigger exited state, intermediate locations trigger waiting state).

Data logging captures comprehensive round information including agent positions, state classifications, group messages, individual decisions, coordinator influence events, and API interaction statistics. The system maintains both in-memory state for active simulation processing and persistent JSON output for subsequent analysis, with intermediate result saving to prevent data loss during extended simulation runs.

The simulation continues until all agents reach exited state or a predefined maximum round limit is exceeded, providing natural termination conditions for both successful evacuation scenarios and extended simulation studies.

\begin{algorithm}
\caption{Commencement Emergency Response Simulation System (Part 3a: Movement \& Speed)}
\begin{algorithmic}[1]

\Procedure{ExecuteMovement}{$\textit{moving\_agents}$, $\textit{density\_map}$}
    \For{$\textit{agent}$ in $\textit{moving\_agents}$}
        \State $\textit{nearby\_count} \leftarrow \textit{density\_map}[\textit{agent.id}]$
        \State $\textit{speed} \leftarrow$ CalculateAdjustedSpeed($\textit{agent}$, $\textit{nearby\_count}$)
        \State $\textit{new\_position} \leftarrow$ PathfindToTarget($\textit{agent.position}$, $\textit{agent.target}$, $\textit{speed}$)
        \State $\textit{agent.position} \leftarrow \textit{new\_position}$
    \EndFor
\EndProcedure

\Procedure{CalculateAdjustedSpeed}{$\textit{agent}$, $\textit{nearby\_count}$}
    \If{$\textit{agent.accessibility} = \text{true}$}
        \State $\textit{base\_speed} \leftarrow 16$, $\textit{min\_speed} \leftarrow 4.8$
    \Else
        \State $\textit{base\_speed} \leftarrow 24$, $\textit{min\_speed} \leftarrow 6.4$
    \EndIf
    \If{CoordinatorNearby($\textit{agent}$)}
        \State $\textit{base\_speed} \leftarrow \textit{base\_speed} \times 1.33$
    \EndIf
    \If{$\textit{nearby\_count} \leq 4$}
        \State $\textit{speed\_factor} \leftarrow 1.0$
    \ElsIf{$\textit{nearby\_count} \geq 30$}
        \State $\textit{speed\_factor} \leftarrow 0.0$
    \Else
        \State $\textit{speed\_factor} \leftarrow 1.0 - \frac{\textit{nearby\_count} - 4}{30 - 4}$
    \EndIf
    \State $\textit{adjusted\_speed} \leftarrow \max(\textit{min\_speed}, \textit{base\_speed} \times \textit{speed\_factor})$
    \State \Return $\textit{adjusted\_speed}$
\EndProcedure

\end{algorithmic}
\end{algorithm}

\begin{algorithm}
\caption{Commencement Emergency Response Simulation System (Part 3b: Pathfinding \& Arrivals)}
\begin{algorithmic}[1]

\Procedure{PathfindToTarget}{$\textit{current}$, $\textit{target}$, $\textit{speed}$}
    \State $\textit{direction} \leftarrow$ UnitVector($\textit{target} - \textit{current}$)
    \State $\textit{intended} \leftarrow \textit{current} + \textit{direction} \times \textit{speed}$
    \State $\textit{obstacles} \leftarrow$ GetStadiumObstacles()
    \For{$\textit{obstacle}$ in $\textit{obstacles}$}
        \State $\textit{intersection} \leftarrow$ LineRectangleIntersection($\textit{current}$, $\textit{intended}$, $\textit{obstacle}$)
        \If{$\textit{intersection} \neq \text{null}$}
            \State $\textit{intended} \leftarrow$ ApplyObstacleSliding($\textit{intersection}$, $\textit{obstacle}$, $\textit{speed}$)
        \EndIf
    \EndFor
    \State \Return ClampToCanvas($\textit{intended}$)
\EndProcedure

\Procedure{CheckDestinationArrivals}{$\textit{moving\_agents}$}
    \For{$\textit{agent}$ in $\textit{moving\_agents}$}
        \If{Distance($\textit{agent.position}$, $\textit{agent.target}$) $\leq 50$}
            \If{$\textit{agent.destination}$ is Exit}
                \State $\textit{agent.state} \leftarrow \text{EXITED}$
            \Else
                \State $\textit{agent.state} \leftarrow \text{WAITING}$
                \State RecordGroupArrival($\textit{agent.group}$, $\textit{agent.destination}$, $\textit{agent.id}$)
            \EndIf
        \EndIf
    \EndFor
\EndProcedure

\Procedure{CheckGroupArrivals}{$\textit{agents}$}
    \State $\textit{resumed} \leftarrow \emptyset$
    \For{$\textit{group}$ in GetAllGroups($\textit{agents}$)}
        \For{$\textit{destination}$ in GetGroupDestinations($\textit{group}$)}
            \State $\textit{chosen} \leftarrow$ GetAgentsWhoChose($\textit{group}$, $\textit{destination}$)
            \State $\textit{arrived} \leftarrow$ GetAgentsWhoArrived($\textit{group}$, $\textit{destination}$)
            \If{$\textit{chosen} = \textit{arrived} \land |\textit{chosen}| > 0$}
                \For{$\textit{agent\_id}$ in $\textit{arrived}$}
                    \State $\textit{agent} \leftarrow$ GetAgent($\textit{agent\_id}$)
                    \State $\textit{agent.state} \leftarrow \text{DISCUSSING}$
                    \State $\textit{resumed} \leftarrow \textit{resumed} \cup \{\textit{agent}\}$
                \EndFor
                \State ClearDestinationTracking($\textit{group}$, $\textit{destination}$)
            \EndIf
        \EndFor
    \EndFor
    \State \Return $\textit{resumed}$
\EndProcedure

\end{algorithmic}
\end{algorithm}

\end{document}